\documentclass[11pt]{article}

\usepackage[preprint]{acl}

\usepackage{times}
\usepackage{latexsym}

\usepackage[T1]{fontenc}

\usepackage[utf8]{inputenc}

\usepackage{microtype}

\usepackage{inconsolata}

\usepackage{amsmath,amssymb}
\usepackage{booktabs,tabularx,array,multirow}
\usepackage{graphicx}
\usepackage{placeins}
\usepackage{xurl}
\usepackage{listings}
\usepackage{enumitem}
\usepackage{siunitx}
\newcommand{\method}{CrypSteer}
\lstdefinestyle{appendixprompt}{
  basicstyle=\ttfamily\footnotesize,
  commentstyle=\ttfamily\upshape,
  columns=fullflexible,
  keepspaces=true,
  breaklines=true,
  breakatwhitespace=true,
  showstringspaces=false,
  frame=single,
  aboveskip=6pt,
  belowskip=6pt
}

\title{Many Repairs, One Direction:\\ Steering Code Models Toward Safer Cryptographic API Use}

\author{
  \textbf{Tianhe Lu, Sakuna Harinda Jayasundara, Eric Spero,} \\
  \textbf{Ewan Tempero, Robert Biddle, Giovanni Russello} \\
  University of Auckland \\
  Auckland, New Zealand \\
  \texttt{\{tlu202,sjay950\}@aucklanduni.ac.nz} \\
  \texttt{\{eric.spero,e.tempero,robert.biddle,g.russello\}@auckland.ac.nz}
}

\begin{document}
\maketitle
\begin{abstract}
Cryptographic API misuse in large language model-generated code can compromise software security.
While many existing approaches improve code security by modifying model parameters or inputs, how representation changes associated with different cryptographic API repairs can support a shared intervention for safer code generation remains underexplored.
To investigate this, we construct 48,000 matched program pairs across four programming languages and ten cryptographic API misuse categories.
Each pair changes a specified API decision while preserving the programming task.
Subtracting the misuse program's internal representation from its secure counterpart's gives a repair vector. 
Across five models, comparisons of mean repair vectors reveal stronger organization by misuse category than by language. 
These category-specific patterns also support a shared intervention.
We introduce \method{}, which extracts a shared direction by normalizing repair differences and selecting a principal component aligned with repairs across misuse categories. 
With model weights fixed, adding one shared direction per model to its internal activations during generation improves aggregate target security rates by 5.28--15.00 percentage points over unsteered generation across languages and misuse categories.
These findings connect the shared structure of cryptographic repair representations to a reusable intervention for safer code generation.



\end{abstract}

\section{Introduction} \label{sec:introduction}


Large language models (LLMs) assist developers with code generation, but they also introduce security vulnerabilities in generated programs~\cite{pearce_2025_asleep, perry_2023_do}.
Software applications rely on cryptographic application programming interfaces (APIs) to protect data, and their secure use requires appropriate algorithms, parameters, and randomness~\cite{kruger_2018_crysl}.
These choices are difficult for developers~\cite{nadi_2016_jumping}, and LLM-generated programs also contain cryptographic API misuses~\cite{Mousavi_2024_investigation}.

\begin{figure*}[t]
\centering
\includegraphics[width=\textwidth]{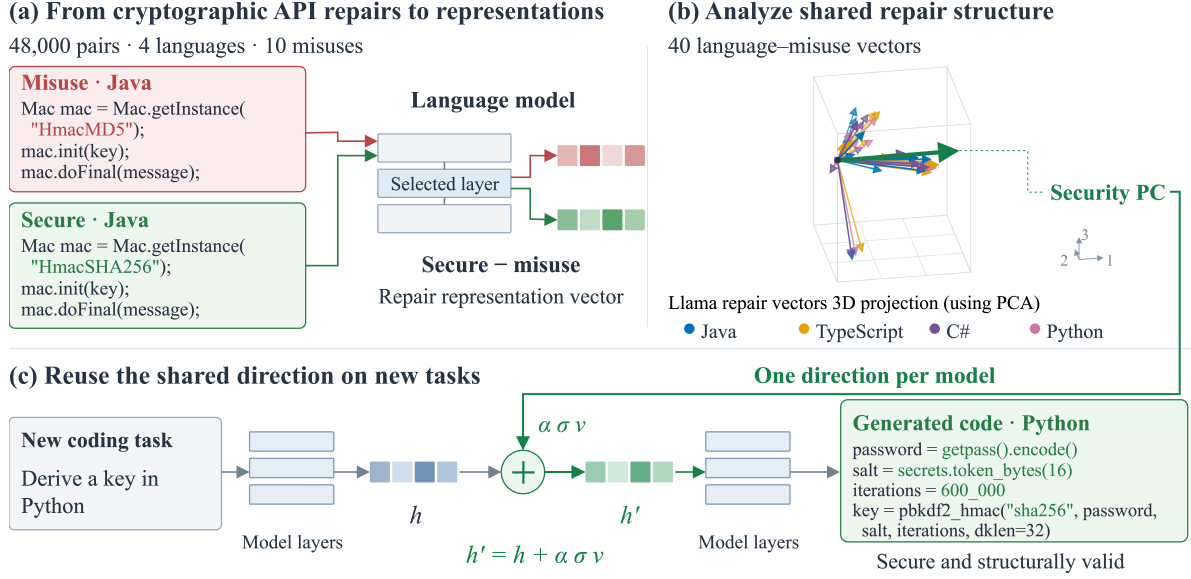}
\caption{Matched repairs support representation analysis and \method{}'s shared generation intervention. 
Secure-minus-misuse program representations provide repair vectors for analysis and direction construction.
The three-dimensional inset visualizes Llama's 40 repair directions for each language-misuse category combination, using uncentered principal component analysis (PCA).
\method{} selects a repair-aligned principal component (security PC) and adds a scaled version to token activations during generation, reusing one direction per model across languages and misuse categories. 
}
\label{fig:study-overview}
\end{figure*}

Existing approaches improve the security of LLM-generated programs through fine-tuning~\cite{he_2023_large, he_2024_instruction}, prompt engineering~\cite{tony_2025_prompting}, retrieval-augmented generation (RAG)~\cite{zhang_2024_seccoder}, and feedback repair~\cite{le_2024_indict}.
However, these gains leave open how models represent security-sensitive choices and whether the associated representation changes can guide generation. 
Activation steering modifies a model's internal activations during generation while keeping model weights and prompts fixed~\cite{rimsky_2024_steering, li_2023_inference, zou_2025_representation, turner_2024_steering, hernandez_2024_inspecting}.
This allows us to study how representations associated with different API misuses can be used to influence the API choices in newly generated programs.


Recent studies show that activation steering directions constructed for individual Common Weakness Enumeration (CWE) categories can reduce vulnerabilities in generated programs~\cite{yan_2026_interpreting, sandoval_2026_surgical, yu_2025_mixture}.
However, different uses of cryptographic APIs involve different security requirements. 
Replacing an unsuitable source of randomness, increasing a key length, and changing an algorithm configuration require different code changes~\cite{Mousavi_2024_investigation, kruger_2018_crysl}.
We investigate how their representation changes vary across languages and misuse categories, and whether these different changes can support one reusable generation intervention.

In this study, we use matched secure and misuse program pairs that  differ only in a specified cryptographic API decision for the same task. 
Repairs may alter an algorithm, a parameter value, or a source of randomness, providing distinct edits whose representation changes can be compared~\cite{Mousavi_2024_investigation, kruger_2018_crysl}.
We construct 48,000 matched secure/misuse program pairs across four programming languages, ten misuse categories, and four code structures. 
We use \emph{repair} to describe the change from a misuse program to its secure counterpart, and use \emph{repair vector} to describe the difference between their program representations. 
By matching the task and surrounding implementation, each vector can be connected to a specified repair. 
Varying language, misuse category, and program structure then allows us to compare these changes across controlled sources of variation.

Across five models, repair vectors are strongly grouped by misuse category, and their shared changes support safer code generation. 
We introduce \method{}, which normalizes repair vectors and selects a component aligned across misuse categories. 
Using \method{} in activation steering improves target security in generated programs.
Comparisons show consistent gains from normalization, with the additional benefit of component selection. 

Figure~\ref{fig:study-overview} summarizes our study. 
We first construct matched secure/misuse programs to support comparisons of repair vectors between them across languages, misuse categories, and code structures (Section~\ref{sec:investigating_cryptographic_repairs_with_matched_programs}).
We then examine how these repair vectors are organized, construct a shared direction (Section~\ref{sec:misuse_category_structure_and_shared_repair_directions}), and test its effects on generation (Section~\ref{sec:steering_code_generation_with_shared_repairs}). 
This paper has the following contributions.
\begin{itemize}[nosep]
    \item \textbf{A controlled resource for studying repairs.} 
    A collection of 48,000 secure/misuse program pairs that contain matched API edits and structural variants enables comparisons of representation changes across programming languages, misuse categories, and program structures. 
    \item \textbf{Evidence connecting repair structure to shared control.} 
    Across five models, repairs differ by misuse category, but they support a shared generation intervention, including when the target misuse category is excluded from direction fitting.
    \item \textbf{A shared steering method for safer generation.} \method{}, which normalizes repair vectors and selects a principal component (PC) aligned across misuse categories, improves aggregate target security in generated programs across all five models using one shared direction per model. 
\end{itemize}

\section{Investigating Cryptographic Repairs with Matched Programs} \label{sec:investigating_cryptographic_repairs_with_matched_programs}
To understand how cryptographic API repairs are represented in models and analyze their organization and causal influence on program generation, we first construct paired programs that isolate a target API decision and use their representation differences to describe repairs. 
For each model, we then choose a common layer at which to analyze those repair vectors and apply generation interventions.

\subsection{Matched Programs with Structural Variants} \label{sec:matched_programs_with_structural_variants}

We build a dataset that keeps the programming task and surrounding implementation scaffolds fixed while changing the target cryptographic API decision for each misuse category listed in Table~\ref{tab:repair-categories}. 
This resource covers Java, TypeScript, C\#, and Python, with ten misuse categories adapted from~\citet{Mousavi_2024_investigation}.
We refer to each language--misuse category combination as a \emph{cell}, e.g., Python--M09.
We use this term for the same grouping throughout the paper.

We design language-specific tasks adapted from~\citet{Mousavi_2024_investigation} and use five GPT models to generate candidate \emph{original} programs with an editable target decision. 
We retain 300 originals in each of the 40 cells, giving 12,000 originals and equal representation for every language and misuse category. 
Based on the original programs and the category-specific contrasts in Table~\ref{tab:repair-categories}, GPT-5.6-sol produces matched secure and misuse versions from each original. 
To examine how code structure affects repair representations, we construct four structural variants of each pair, which forms across two choices. 
This includes whether the target value reaches the API through a named variable or directly at the call site, and whether it is constructed inline or returned by a helper function.
Both programs in a pair use the same structural form, and each original contributes four matched pairs, or eight programs. 
Therefore, the full resource contains 48,000 pairs, or 96,000 programs. 
We conduct a quality check of the resource that covers required operations, arguments, dataflow, and target labels, which we specify in Appendix~\ref{app:generation_rewriting_and_validation}.
The security rules follow category-specific policies from~\citet{Mousavi_2024_investigation}.
Appendix~\ref{app:resource_construction} gives the construction procedure. 

\begin{table}[tbp]
\centering
\footnotesize
\setlength{\tabcolsep}{1pt}
\begin{tabularx}{\columnwidth}{@{}l>{\raggedright\arraybackslash}X>{\raggedright\arraybackslash}X@{}}
\toprule
ID & Misuse definition & Dataset contrast  (secure / misuse) \\
\midrule
M01 & Predictable cryptographic keys & Random / fixed literal key \\
M02 & Insecure mode of encryption & CBC mode / ECB mode \\
M03 & Predictable IVs in CBC mode & Random / constant IV \\
M04 & Short cryptographic keys & 2048 / 1024 bits for RSA \\
M05 & Predictable random output & Cryptographic / general-purpose PRNG \\
M06 & Predictable salts for key derivation & Random / constant salt \\
M07 & Low iteration counts for key derivation & 100,000 / 100 iterations \\
M08 & Hardcoded constant passwords & Environment / hardcoded password \\
M09 & Broken hash functions & SHA-256 / SHA-1 \\
M10 & Compromised MAC algorithms & HMAC-SHA-256 / HMAC-MD5 \\
\bottomrule
\end{tabularx}
\caption{Misuse categories adapted from~\citet{Mousavi_2024_investigation} and instantiated in four languages. 
Each contrast lists the secure and misuse implementations used to construct a matched pair. 
CBC and ECB denote cipher block chaining and electronic codebook modes, respectively. 
IV denotes initialization vector.
PRNG denotes pseudorandom number generator.
MAC denotes message authentication code.
Appendix~\ref{app:security_of_api_usage} specifies the label policies.}
\label{tab:repair-categories}
\end{table}

\subsection{Program Representation and Repair Vectors} \label{sec:program_representation_and_repair_vectors}

We supply each program's complete source code $x$ directly to the model in one forward pass.
At layer $\ell$ and token position $t$, we record \emph{token activations} $h_{\ell,t}(x)$ from the residual stream, which carries the current token representations between decoder blocks. 
Averaging these activations over the program body gives one fixed-size \emph{program representation}, $\bar h_\ell(x)$, following mean-pooling approaches to code representation~\cite{jain_2021_contrastive, guo_2022_unixcoder}.
The leading imports and package declarations remain in the input as context but are excluded from the average, so the program representation summarizes the implementation, including definitions, API calls, and later uses. 

For original $i$ in structural form $s$ at layer $\ell$, let $\bar h^{\mathrm{secure}}_{i,s,\ell}$ and $\bar h^{\mathrm{misuse}}_{i,s,\ell}$ be the paired program representations.
Their secure-minus-misuse difference defines a \emph{structure-specific repair vector} $\delta_{i,s,\ell}$, whose orientation and length capture the representation change associated with the specified edit in that structural form.
\begin{equation}
 \delta_{i,s,\ell}=
 \bar h^{\mathrm{secure}}_{i,s,\ell}-\bar h^{\mathrm{misuse}}_{i,s,\ell}.
\end{equation}
We combine the four forms of each original by averaging their structure-specific repair vectors:
\begin{equation}
 d_{i,\ell}=\frac{1}{4}\sum_{s=1}^{4}\delta_{i,s,\ell}.
\label{eq:original-repair}
\end{equation}
This gives one  \emph{original-level repair vector} $d_{i,\ell}$ per original, with equal weights for its four structural forms. 
We find that cross-structure repair vectors have strong direction alignment, which we report in Appendix~\ref{app:representation_stability_across_program_structures}.
This supports using these original-level repair vectors to study category-level patterns. 

Subsequent geometry analysis and direction fitting use these 12,000 vectors  $d_{i,\ell}$, one per original, and we omit $\ell$ for the selected layer. 
Layer selection classifies individual implementations, while intervention scale calibration needs to capture variation across structural forms, so they both use all 96,000 program representations. 


\subsection{Models and Layer Selection} \label{sec:models_and_layer_selection}
We study five instruction-tuned models spanning general-purpose and code-specialized models to understand whether different LLM structure and architecture influence the results. 
The models include: Llama-3.1-8B-Instruct (Llama)~\cite{Aaron_2024_llama}, CodeGemma-7B-IT (CodeGemma)~\cite{codegemmateam_2024_codegemma}, Qwen2.5-7B-Instruct (Qwen)~\cite{qwen_2025_qwen25}, Qwen2.5-Coder-7B-Instruct~(Qwen-Coder) \cite{hui_2024_qwen25}, and DeepSeek-Coder-V2-Lite-Instruct (DeepSeek)~\cite{deepseekai_2024_deepseekcoderv2}.
Each model has its own representation space, selected layer and steering directions. 
We use the short names in parentheses throughout the paper. 


To reuse one direction across languages and misuse categories, we select one layer that contributes most for the cryptographic API misuse concept for representation analysis and generation intervention within each model.
We compare layers by testing how well a linear classifier can distinguish secure from misuse programs using their program representations.
At each layer, we train a logistic regression classifier, called a \emph{linear probe}, to predict secure/misuse labels from the program representations~\cite{belinkov_2022_probing}.
We hold out one misuse category or one language at a time to test the generalizability of the classification, and train a separate probe on the remaining groups and evaluate it on the held-out group.
This results in ten misuse category folds and four language folds at each layer. 

We use the area under the receiver operating characteristic curve (AUROC) to measure the ranking performance of each probe for each fold~\cite{tom_2006_introduction}.
We first average the ten misuse category scores and the four language scores separately, then give the two averages equal weight, to give both transfer equal importance. 
The layer with the highest combined score is selected as the common layer for that model.  
The selected layers lie in the middle to upper-middle parts of the model, consistent with empirical steering studies of alignment~\cite{rimsky_2024_steering}, personality~\cite{feng_2026_persona}, sentiment~\cite{zhao_2026_denoising}, and emotion-driven preferences~\cite{sofroniew_2026_emotion}.
Appendix~\ref{app:linear_probes_and_layer_selection} gives the selected layers and optimization settings.

\begin{figure}[t]
  \centering
  \includegraphics[width=\columnwidth]{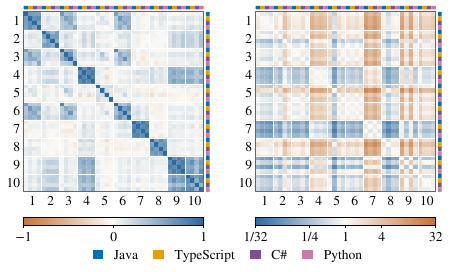}
  \par\nointerlineskip
  \begin{minipage}[t]{0.49\columnwidth}
    \centering
    {\footnotesize (a) Direction cosine\par}
  \end{minipage}\hfill
  \begin{minipage}[t]{0.49\columnwidth}
    \centering
    {\footnotesize (b) Norm ratio\par}
  \end{minipage}
  \caption{Geometry of Llama repair vectors at layer 14. (a) Cosine similarities between 40 cell-mean directions. (b) Row/column ratios of 40 cell-average repair norms. On both axes, 1–10 denote misuse categories M01–M10. Within each category, rows and columns follow the order Java, TypeScript, C\#, and Python.}
  \label{fig:repair-geometry}
\end{figure}

\section{Misuse Category Structure and Shared Repair Directions} \label{sec:misuse_category_structure_and_shared_repair_directions}
Different uses of cryptographic APIs involve different security requirements, but they may produce distinct representation changes while still sharing an orientation that can guide generation.
We examine this possibility by comparing repair vectors across misuse categories and languages, constructing a shared direction, and testing its alignment with new programs and its responses to combined repairs.

\subsection{Representation Organization Across Languages and Misuse Categories} \label{sec:representation_organization_Across_languages_and_misuse_categories}

We use cosine similarity to measure directional alignment of repair vectors across misuse categories and languages~\cite{ethayarajh_2019_contextual}.
Cosine values near one indicate similar orientations, zero indicates orthogonality, and negative values indicate opposing orientations. 
At each model's selected layer, we average the 300 original repair vectors within each language--category cell, which gives 40 cell-mean directions. 
The cosine heatmap in Figure~\ref{fig:repair-geometry}(a) shows the directional structure for Llama, and Table~\ref{table-geometry-summary} compares same-category cosines across languages with different-category cosines within languages for all five models. 
Directions from the same misuse category across languages are substantially more aligned than directions from different misuse categories within one language.

We also test whether a repair category can be identified using directions from other languages. 
For each language, we build ten misuse category references by averaging the unit cell-mean directions from the other three languages. 
Each of its ten cell directions is then assigned to the reference with the highest cosine similarity. 
This gives 40 queries across four languages and ten misuse categories. 
Results in Table~\ref{table-geometry-summary} show that this recovers the correct category in 38--40 of 40 cases per model. 
Together, these results identify misuse category as the stronger organizing factor while retaining language-dependent variation.

\begin{table}[t]
\centering
\scriptsize
\setlength{\tabcolsep}{1pt}
\begin{tabular*}{\columnwidth}{@{\extracolsep{\fill}}lcccc@{}}
\toprule
Model & \shortstack{Mean cosine\\ same category \\across languages} & \shortstack{Mean cosine\\different categories\\within a language} & \shortstack{Category\\matches} & \shortstack{Norm\\ratio}\\
\midrule
Llama & 0.607 & 0.130 & 40/40 & 19.2\\
CodeGemma & 0.673 & 0.111 & 40/40 & 22.5\\
Qwen & 0.673 & 0.095 & 38/40 & 14.6\\
Qwen-Coder & 0.694 & 0.118 & 38/40 & 14.3\\
DeepSeek & 0.650 & 0.112 & 38/40 & 16.2\\
\bottomrule
\end{tabular*}
\caption{Full-corpus repair geometry at the selected layer. Category matches use references from other languages. Norm ratio compares the largest and smallest cell-average repair norms.}
\label{table-geometry-summary}
\end{table}



However, repair magnitudes vary substantially. 
For this comparison, we average the 300 original repair $L_2$ norms within each cell and compare the largest and smallest cell averages. 
A value near one means similar vector lengths. 
Figure~\ref{fig:repair-geometry}(b) shows the cell-wise norm ratio comparison for Llama. 
Table~\ref{table-geometry-summary} shows that the largest cell-averaged repair norm is 14.3--22.5 times the smallest across five models. 
Therefore, simply averaging the raw repair vectors will give some repairs substantially more influence than others. 
This motivates comparing aggregation of the raw repair vectors with aggregation that first gives every repair equal length.

We also test whether the structure of repair representations is stable across layers. 
We compute the pairwise cosine similarities between the repair directions at each layer, and then compute the pairwise cosine similarities of these cosine similarity matrices across layers to analyze the representational similarity~\cite{kriegeskorte_2008_representational}.
We find that the structure of repair geometry is relatively stable throughout most of the model, particularly from early-middle to late layers. 
At the probe-selected layers for each model, similarity to the immediately preceding and following layers ranges from 0.993 to 0.998. 
Therefore, the selected layers lie within regions of stable repair relationships. 
Appendix~\ref{app:additional_geometry_analysis_and_clustering} provides detail of the geometry and additional clustering analyses.

\subsection{Constructing a Shared Direction} \label{sec:constructing_a_shared_direction}
Distinct misuse category repair vectors can align with a common axis.
\method{} seeks such an axis by first normalizing repair vectors to give each repair equal length, then selecting a component with positive mean alignment in as many misuse categories as possible. 
This construction turns the observed misuse category structure into a candidate for shared generation control.

We compare \method{} with three variants of difference-in-means (DiffMean) constructions. 
For matched pairs, averaging secure-minus-misuse repair vectors is equivalent to subtracting the mean misuse representation from the mean secure representation~\cite{rimsky_2024_steering}.
\emph{Cell DiffMean} averages the 300 original repair vectors within each cell, giving 40 directions per model. 
\emph{Pooled DiffMean} averages all repair vectors across 40 cells. 
\emph{Normalized DiffMean} first normalizes each repair vector, then averages across cells.
The latter two each produce one shared direction per model. 
Every steering vector is normalized to unit length before intervention for a fair comparison. 
Appendix~\ref{app:direction_construction} gives the formulas.

For \method{}, let $d_i=d_{i,\ell}$ and $u_i=d_i/\lVert d_i\rVert_2$ at the selected layer, with $N=12,000$ originals.
We apply uncentered singular value decomposition (SVD) to the matrix with rows $u_i$ to approximate the unit repair vectors by their projections onto a common axis. 
Keeping the vectors uncentered preserves the origin as the point of no change between paired program representations. 
SVD finds axes that capture large squared projections of the unit repair vectors. 
The first component solves
\begin{equation}
 v_1\in\arg\max_{\lVert v\rVert_2=1}
       \frac1N\sum_i (u_i^\top v)^2.
 \label{eq:repair-}
\end{equation}
Geometrically, $v_1$ defines the line through the origin that minimizes the sum of squared distances from the unit repair vectors to the line~\cite{eckart_1936_approximation}.
Each repair $u_i$ has an approximation which is the projection $(u_i^\top v_1)v_1$ onto this axis.
Each later component maximizes the same objective subject to orthogonality to all earlier components.

The squared objective treats an axis and its reverse equally, so we use the secure-minus-misuse orientation of the repairs to choose a component and its sign.
We consider both signs of the first ten PCs, which are right singular vectors $v_1,\ldots,v_{10}$.
For each candidate, we average the projections of the repair vectors within each misuse category and count how many category means are positive. 
We select the candidate with the largest count, with detailed rules in Appendix~\ref{app:direction_construction}. 
This criterion prioritizes alignment across misuse categories among the leading components. 
The selected unit direction $v_{\mathrm{CrypSteer}}$, or \emph{security PC}, is PC2 for Qwen and PC1 for the other four models.
\method{} uses this model-specific direction across all four languages and ten misuse categories in activation steering.

\subsection{Alignment on Held-Out Programs} \label{sec:alignment_on_held_out_programs}

\begin{table}[t]
\centering
\footnotesize
\setlength{\tabcolsep}{3pt}
\begin{tabular}{@{}lccc@{}}
\toprule
Model & Positive repairs (\%) & Mean cosine & Positive cells \\
\midrule
Llama & 95.36  & 0.343 & 38/40\\
CodeGemma & 88.86  & 0.332 & 36/40\\
Qwen & 95.22  & 0.324 & 38/40\\
Qwen-Coder & 91.11 & 0.318 & 37/40\\
DeepSeek & 97.86  & 0.337 & 40/40\\
\bottomrule
\end{tabular}
\caption{Held-out alignment of repair vectors. Positive repairs have cosine above zero. Positive cells have positive mean cosine.}
\label{tab:repair-heldout}
\end{table}
\begin{table}[t]
\centering
\footnotesize
\setlength{\tabcolsep}{3pt}
\begin{tabular*}{\columnwidth}{@{\extracolsep{\fill}}lcccc@{}}
\toprule
Model & Java & TypeScript & C\# & Python\\
\midrule
Llama & 0.930 & 0.819 & 0.971 & 0.990 \\
CodeGemma & 0.770 & 0.615 & 0.838 & 0.946 \\
Qwen & 0.986 & 0.996 & 0.997 & 0.987 \\
Qwen-Coder & 0.950 & 0.964 & 0.866 & 0.986 \\
DeepSeek & 0.978 & 0.991 & 0.985 & 0.965 \\
\bottomrule
\end{tabular*}
\caption{Prediction $R^2$ for 1,013 multi-repair combinations per language using the all-misuse and ten single-repair programs as references.}
\label{tab:combination-prediction}
\end{table}

We test whether the shared direction aligns with repairs excluded from its construction. 
With the layer selected from the full corpus, we split originals into 70\% discovery and 30\% test sets, stratified by language, misuse category, and task.
The split contains 8,400 discovery originals and 3,600 test originals, with all four structural variants of an original kept together. 
Using only original-level repair vectors in the discovery set, we perform SVD and select the component and its orientation.

We keep the resulting direction fixed and compute its cosine similarity with each test repair vector. 
Table~\ref{tab:repair-heldout} reports the proportion of positive cosines, the mean cosine over all test originals, and the number of cells with positive mean cosine.  
Positive alignment holds for 88.86\%--97.86\% of test originals, with positive mean test cosine in 36--40 cells per model.
Therefore, the shared orientation extends to repairs beyond the direction-fitting set. 



\begin{figure}[t]
  \centering
  \includegraphics[width=\columnwidth]{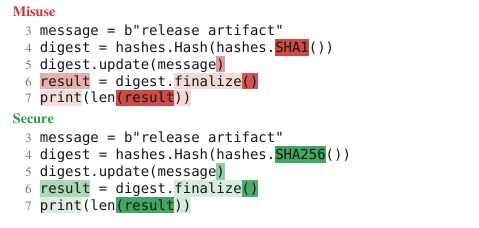}
  \caption{Python M09 paired span projections using the Llama layer-14 \method{} direction. Green and red indicate positive and negative paired projection differences.}
  \label{fig:repair-heatmaps}
\end{figure}

\subsection{Response to Individual and Combined Security Choices} \label{sec:response_to_individual_and_combined_security_choices}

We use additional controlled programs to test how responses to individual repairs combine within a program.
Each language has one program template with ten operations corresponding to the ten misuse categories, each switchable between secure and misuse implementations. 
This gives $2^{10}=1,024$ combinations per language that separate from the direction-fitting resource. 
We measure each program's scalar projection, which is the dot product of its representation with the fixed unit direction.

The all-misuse program and the ten single-repair programs provide eleven reference measurements. 
For each repair, we measure the projection change from the all-misuse program.
We predict a program with several repairs by adding their individual changes to the all-misuse projection, then evaluate these predictions on the remaining 1,013 combinations. 
Table~\ref{tab:combination-prediction} reports the prediction $R^2$, which measures prediction accuracy.
Across the five models and four languages, the median is 0.968.
Therefore, individual repair responses often combine approximately additively. 
Appendix~\ref{app:controlled_program_diagnostics} gives the prediction rule and a complementary descriptive analysis that fits an additive model to all 1,024 configurations.

To examine where a repair changes the representation of a program, we also curate paired programs with secure and misuse decisions for each language and misuse category.
We use paired heatmaps to show how individual repairs affect token activation projections onto the fixed shared direction, which shows their direction similarity.
In Figure~\ref{fig:repair-heatmaps}, replacing SHA-1 with SHA-256 changes projections at the algorithm choice and the unchanged \texttt{result} in \texttt{result=digest.finalize()} and \texttt{print(len(result))}.
The unchanged expressions receive different activations because their preceding context has changed. 
This connects the API repair to the model's representation of a later output expression. 
Appendix~\ref{app:paired_code_heatmaps} gives the heatmap construction procedure. 


\section{Steering Code Generation with Shared Repairs} \label{sec:steering_code_generation_with_shared_repairs}
\begin{table*}[tp]
\centering
\footnotesize
\setlength{\tabcolsep}{1pt}
\begin{tabular*}{\textwidth}{@{\extracolsep{\fill}}llc*{2}{cccc}@{}}
\toprule
\multirow{2}{*}{Model} & \multirow{2}{*}{Metric (\%)} &
\multicolumn{1}{c}{$\alpha=0$} &
\multicolumn{4}{c}{$\alpha=-1$} &
\multicolumn{4}{c}{$\alpha=+1$} \\
\cmidrule(lr){3-3}\cmidrule(lr){4-7}\cmidrule(l){8-11}
& & \multicolumn{1}{c}{Baseline} &
\shortstack{Cell\\DiffMean} & \shortstack{Pooled\\DiffMean} &
\shortstack{Normalized\\DiffMean} & \method{} &
\shortstack{Cell\\DiffMean} & \shortstack{Pooled\\DiffMean} &
\shortstack{Normalized\\DiffMean} & \method{} \\
\midrule
\multirow{3}{*}{Llama} & \textnormal{secure@1} & 62.08 & 39.28 & 45.97 & 37.58 & \underline{33.22} & 71.90 & 72.12 & 74.88 & \textbf{77.08} \\
 & \textnormal{valid@1} & 72.67 & 66.97 & 64.73 & 58.67 & 53.78 & 68.23 & 70.02 & 69.47 & 69.50 \\
 & \textnormal{secure-valid@1} & 47.52 & 28.25 & 32.12 & 23.08 & 18.20 & 52.33 & 54.40 & 56.33 & \textbf{57.18} \\
\midrule
\multirow{3}{*}{CodeGemma} & \textnormal{secure@1} & 46.27 & 32.23 & 33.03 & 30.57 & \underline{28.63} & 54.07 & 53.73 & 55.45 & \textbf{56.68} \\
 & \textnormal{valid@1} & 37.92 & 36.27 & 38.30 & 36.47 & 35.25 & 39.43 & 36.95 & 38.37 & 39.30 \\
 & \textnormal{secure-valid@1} & 21.33 & 12.63 & 15.28 & 13.37 & 12.22 & 25.67 & 24.30 & 25.95 & \textbf{26.53} \\
\midrule
\multirow{3}{*}{Qwen} & \textnormal{secure@1} & 56.17 & 38.95 & 41.55 & \underline{35.90} & 36.15 & 63.65 & 64.10 & \textbf{67.95} & 66.97 \\
 & \textnormal{valid@1} & 74.68 & 71.55 & 71.20 & 68.62 & 69.70 & 70.77 & 72.03 & 73.10 & 72.00 \\
 & \textnormal{secure-valid@1} & 44.83 & 29.63 & 32.52 & 27.07 & 27.38 & 48.67 & 48.65 & \textbf{51.58} & 50.53 \\
\midrule
\multirow{3}{*}{Qwen-Coder} & \textnormal{secure@1} & 64.23 & 54.85 & 55.32 & 51.47 & \underline{50.43} & 67.77 & 68.50 & \textbf{69.70} & 69.52 \\
 & \textnormal{valid@1} & 73.82 & 74.87 & 74.77 & 74.80 & 74.82 & 72.68 & 71.88 & 71.23 & 72.15 \\
 & \textnormal{secure-valid@1} & 50.58 & 44.62 & 45.23 & 42.67 & 41.35 & 51.93 & 51.40 & 52.28 & \textbf{52.90} \\
\midrule
\multirow{3}{*}{DeepSeek} & \textnormal{secure@1} & 61.87 & \underline{50.20} & 54.50 & 50.38 & 50.82 & \textbf{69.03} & 67.62 & 68.47 & 68.72 \\
 & \textnormal{valid@1} & 89.12 & 88.17 & 87.17 & 87.28 & 87.37 & 86.40 & 86.92 & 86.67 & 86.92 \\
 & \textnormal{secure-valid@1} & 55.83 & 44.57 & 48.25 & 44.68 & 45.05 & 60.80 & 59.98 & 60.62 & \textbf{60.87} \\
\bottomrule
\end{tabular*}
\caption{
Single-generation success rates over 6,000 outputs per model and condition.
Bold marks the highest \texttt{secure@1} and \texttt{secure-valid@1} among the four directions at $\alpha=+1$ for each model.
Underlining marks the lowest \texttt{secure@1} at $\alpha=-1$, corresponding to the largest observed decrease.}
\label{tab:generation-main}
\end{table*}

The representation analyses show that repairs have category-specific structure while aligning with a shared direction. 
We then test whether that direction can guide security-sensitive choices when a model generates a new program.
The experiments compare repair aggregation methods, vary the sources used to fit a direction, and examine the effect of intervention strength.


\subsection{Intervention and Evaluation} \label{sec:intervention_and_evaluation}

We steer model generation by adding a scaled unit direction to the selected layer's residual stream activation at the final prompt position and each generation step:
\begin{equation}
 h'_{t}=h_{t}+\alpha\sigma v,
 \label{eq:generation-intervention}
\end{equation}
where $h_{t}$ is the token activation at position $t$ at the selected layer, $v$ is a unit direction, and $\alpha$ controls the sign and strength of the intervention.

We measure the intervention strength in units of $\sigma$, which is the square root of the total sample variance of all program representations at the selected layer.
This measures how widely program representations are spread around their mean, and provides a model-specific reference scale for intervention. 
Because $v$ has a unit length, the update in Equation~\ref{eq:generation-intervention} has length $|\alpha|\sigma$.

Each condition uses five fixed generation prompts per language--misuse category cell and 30 generation attempts per prompt, giving 200 prompts and 6,000 outputs per model. 
These prompts are designed to form a separate catalog covering cryptographic functionalities shared with the resource construction tasks. 
We reuse the prompts and attempt seeds across conditions.
We evaluate Cell DiffMean, Pooled DiffMean, Normalized DiffMean, and \method{} at $\alpha=\pm1$, and additionally test \method{} at $0.5\sigma$ and $2\sigma$.
Appendix~\ref{app:scale_and_generation_settings} gives the scale formula and generation settings.

We measure three single-sample success rates using the $k=1$ case of the \texttt{pass@k} estimator~\cite{chen_2021_evaluating}.
\texttt{secure@1} is the proportion of outputs labeled secure under the target cryptographic API usage policy, \texttt{valid@1} is the proportion that pass structural validity checks, and \texttt{secure-valid@1} requires both on the same output. 
Structural validity checks cover source-level parsing or compilation, imports, and required API operations, arguments, and connections. 
Appendix~\ref{app:evaluation_and_output_audits} reports a blinded manual review of evaluator labels. 

For each prompt, we calculate the fraction of its 30 attempts that satisfies each criterion, retaining all attempts in the denominator.
Aggregate results give equal weight to each of the 40 cells. 
Table~\ref{tab:generation-main} reports the main results, and Appendix~\ref{app:generation_evaluation_details_and_subgroup_results} gives paired statistics and results for each language and misuse category.

\subsection{Shared Control and Aggregation Choices} \label{sec:shared_control_and_aggregation_choices}

Using one shared direction per model for all four languages and ten misuse categories, \method{} improves the security of generated programs for all five models. 
As shown in Table~\ref{tab:generation-main}, at $+1\sigma$, \method{} improves \texttt{secure@1} by 5.28--15.00 percentage points (pp) and \texttt{secure-valid@1} by 2.32--9.67 pp.
Reversing the direction also reduces aggregate security.
These opposing effects provide causal evidence that the direction is a shared representation of cryptographic API misuse that can be reused for generation-time intervention. 
\method{} even achieves a higher observed \texttt{secure-valid@1} than Cell DiffMean in all five models, and Appendix~\ref{app:generation_evaluation_details_and_subgroup_results} shows this also holds for most language and misuse category groups.
This suggests that aggregating repairs across tasks could be more useful than restricting construction to closely matched examples. 
A possible explanation is that repair differences reflect both the target security change and its interaction with the surrounding implementation.
Aggregation may emphasize recurring security-related changes while reducing the influence of components specific to other contexts.

In addition to \method{}, Pooled DiffMean and Normalized DiffMean also improve observed \texttt{secure@1} and \texttt{secure-valid@1} over the unsteered baseline.
Normalizing repairs before averaging consistently improves both rates over Pooled DiffMean. 
Raw averaging gives longer repair vectors larger influence, while Normalized DiffMean gives each repair equal weight regardless of length. 
Together with the norm variation analysis, this comparison supports combining orientations with equal influence for each repair. 
Component selection gives \method{} the highest \texttt{secure-valid@1} improvement in four models. 
This suggests that component selection captures more distribution structure of repair vector directions and the cross-misuse category alignments, and provides additional security benefits.

\subsection{Reuse with Excluded Fitting Sources} \label{sec:reuse_with_excluded_fitting_sources}

\begin{table}[tbp]
\centering
\footnotesize
\setlength{\tabcolsep}{1pt}
\begin{tabular*}{\columnwidth}{
  @{\extracolsep{\fill}}
  ll
  *{4}{S[table-format=2.2, table-number-alignment=center]}
  @{}
}
\toprule
 &  & \multicolumn{4}{c}{Excluded from direction fitting} \\
\cmidrule(l){3-6}
Model & Metric & {None} & {Language} & {Misuse} & {Both} \\
\midrule
\multirow{2}{*}{Llama} & secure@1 & 15.00 & 15.00 & 13.20 & 13.42 \\
 & secure-valid@1 & 9.67 & 9.80 & 8.57 & 9.05 \\
\addlinespace[2pt]
\multirow{2}{*}{\shortstack[l]{Code\\Gemma}} & secure@1 & 10.42 & 9.53 & 8.27 & 8.02 \\
 & secure-valid@1 & 5.20 & 5.05 & 4.40 & 4.50 \\
\addlinespace[2pt]
\multirow{2}{*}{Qwen} & secure@1 & 10.80 & 11.37 & 10.20 & 9.88 \\
 & secure-valid@1 & 5.70 & 5.98 & 5.65 & 5.60 \\
\addlinespace[2pt]
\multirow{2}{*}{\shortstack[l]{Qwen-\\Coder}} & secure@1 & 5.28 & 5.02 & 3.87 & 4.47 \\
 & secure-valid@1 & 2.32 & 1.77 & 1.57 & 1.62 \\
\addlinespace[2pt]
\multirow{2}{*}{DeepSeek} & secure@1 & 6.85 & 5.10 & 6.53 & 6.73 \\
 & secure-valid@1 & 5.03 & 2.82 & 4.18 & 4.58 \\
\bottomrule
\end{tabular*}
\caption{\method{} gains over the unsteered baseline in percentage points at $+1\sigma$. None denotes full-source fitting.}
\label{tab:generation-source-exclusion}
\end{table}
\begin{table}[t]
\centering
\footnotesize
\setlength{\tabcolsep}{1.5pt}
\begin{tabular}{@{}llccc@{}}
\toprule
Model & Method & \texttt{secure@1} & \texttt{valid@1} &
\shortstack{\texttt{secure-}\\\texttt{valid@1}} \\
\midrule
\multirow{3}{*}{Llama}
 & SCS-Code & 75.45 & 72.62 & \textbf{58.47} \\
 & Surgical Repair & 66.57 & 70.87 & 50.62 \\
 & \method{} & \textbf{77.08} & 69.50 & 57.18 \\
\midrule
\multirow{3}{*}{CodeGemma}
 & SCS-Code & 51.50 & 40.35 & 25.02 \\
 & Surgical Repair & 51.20 & 39.58 & 23.80 \\
 & \method{} & \textbf{56.68} & 39.30 & \textbf{26.53} \\
\midrule
\multirow{3}{*}{Qwen}
 & SCS-Code & 63.17 & 73.93 & 50.05 \\
 & Surgical Repair & 60.30 & 71.42 & 45.12 \\
 & \method{} & \textbf{66.97} & 72.00 & \textbf{50.53} \\
\midrule
\multirow{3}{*}{Qwen-Coder}
 & SCS-Code & 65.40 & 74.30 & 51.43 \\
 & Surgical Repair & 64.43 & 74.27 & 50.95 \\
 & \method{} & \textbf{69.52} & 72.15 & \textbf{52.90} \\
\midrule
\multirow{3}{*}{DeepSeek}
 & SCS-Code & 62.15 & 85.98 & 54.98 \\
 & Surgical Repair & 63.40 & 88.52 & 56.63 \\
 & \method{} & \textbf{68.72} & 86.92 & \textbf{60.87} \\
\bottomrule
\end{tabular}
\caption{Direction construction comparisons under common $+1\sigma$ steering. Bold marks the highest \texttt{secure@1} and \texttt{secure-valid@1} among the three methods.}
\label{tab:generation-literature}
\end{table}

We next test how much shared control depends on including the target language or misuse category in direction fitting. 
For each target cell, language exclusion retains all ten misuse categories in the other three languages. 
Misuse category exclusion retains the other nine misuse categories in all four languages. 
Joint exclusion retains the other nine misuse categories in the other three languages. 
These fits use 9,000, 10,800, and 8,100 original-level repair vectors, respectively.



We repeat SVD, component and orientation selection on each retained subset.
A language-excluded direction is reused for that language's ten misuse categories.
A misuse category-excluded direction is reused for that misuse category in four languages. 
A joint exclusion direction is used for the target cell. 
This gives four, ten, and forty directions per model.
We evaluate the resulting directions on the same task descriptions, with the same layer and scale determined from the full corpus.

As shown in Table~\ref{tab:generation-source-exclusion}, \method{}'s useful control extends beyond the languages and misuse categories used to fit a direction. 
All exclusion settings retain positive security gains for every model, supporting reuse across languages and misuse categories. 
This provides an interpretation for repair vector geometry.
Different misuse categories have different repair patterns, but changes from retained misuse categories can guide safer choices in the excluded misuse category. 


\subsection{Comparing with Other Security Steering Methods} \label{sec:comparing_with_other_security_steering_methods}

We compare \method{} with two other ways to obtain a steering direction. 
We use SCS-Code's code-choice judgments~\cite{wendlinger_2026_security} and Surgical Repair's contrasting implementation instructions~\cite{sandoval_2026_surgical}.
We use our paired program repair dataset for SCS-Code, and design a set of contrasting instructions for uses of cryptographic APIs for Surgical Repair. 
We extract the activation differences following their respective extraction procedures, and average them to derive a shared direction for activation steering. 
We normalize each vector to obtain a steering direction and evaluate it with the same prompts, layer, scale, and $+1\sigma$ intervention.

Table~\ref{tab:generation-literature} shows that \method{} achieves the highest security rate in all models, while the joint outcome depends on how well the intervention preserves structural validity. 
For Llama, SCS-Code yields more structurally valid outputs and a slightly higher \texttt{secure-valid@1}.
The results suggest that program repair differences are a useful source of steering directions, while different methods have different trade-offs between target security and structural validity. 
Appendix~\ref{app:comparison_with_scscode_and_surgical_repair} details the input construction, extraction, and paired comparisons.

\subsection{Intervention Strength and Outcome Variation} \label{sec:intervention_strength_and_outcome_variation}

\begin{figure}[t]
\centering
\includegraphics[width=\columnwidth]{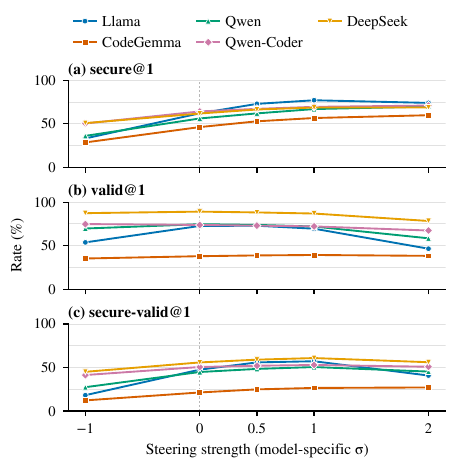}
\caption{\method{} strength responses, with 6,000 outputs per point. Zero is the unsteered baseline.}
\label{fig:generation-strength}
\end{figure}

Steering strength affects the balance between target security and structural validity.
As shown in Figure~\ref{fig:generation-strength}, moderate positive steering improves security and the joint outcome together. 
But moving from $+1\sigma$ to $+2\sigma$ reduces structural validity and lowers \texttt{secure-valid@1} in four models. 
The curves support $+1\sigma$ as a practical reference strength across the five models. 
Scaling the intervention by each model's program representation spread gives a useful balance between security and validity across the evaluated models.

This distinction between target security and structural validity parallels the separation between preference and utility in~\citet{xu_2026_steering}.
Here, the target preference is a secure API choice, and utility concerns whether the implementation satisfies the requested program requirements measured by structural validity. 
These objectives interact because a changed API choice must remain compatible with imports, arguments, and operations. 
Their interaction motivates a complementary direction learned from repairs that restore structural validity while retaining the target security choice.

\section{Related Work} \label{sec:related_work}

\paragraph{Improving the Security of LLM-Generated Code.} \label{sec:improving_the_security_of_LLM_generated_code}
Researchers have developed several techniques for improving the security of LLM-generated code. 
Training-based approaches use security-related examples or objectives to update model parameters or learn additional prefixes~\cite{he_2023_large, he_2024_instruction, xu_2025_prosec}.
Prompt-based approaches provide explicit security instructions and optimize prompt wording~\cite{tony_2025_prompting, nazzal_2024_promsec}.
Retrieval-based methods supply security-related knowledge, examples, and coding guidelines in the input context~\cite{zhang_2024_seccoder, tony_2025_retrieve, lu_2026_retrieved}. 
Decoding-time control methods guide next-token selection using a security-focused helper model or masking rule~\cite{li_2024_cosec, storhaug_2023_efficient}.
Feedback-guided generation and repairs use program analysis results to revise the candidate program~\cite{le_2024_indict, kulsum_2024_case}.
These methods can also be combined, and a system can first retrieve security guidelines and then iteratively revise the generated program~\cite{tony_2025_retrieve}.
Our study connects controlled API edits to internal representation changes and generation-time intervention for safer code generation.

\paragraph{Activation Steering.} \label{sec:activation_steering}
Activation steering is a control technique within representation engineering that guides generation by modifying a model's internal activations~\cite{subramani_2022_extracting, rimsky_2024_steering}.
Subtracting the average activations of negative examples from those of positive examples gives a steering direction known as DiffMean~\cite{rimsky_2024_steering}.
The DiffMean direction has outperformed classifier-weight directions in evaluations of truthfulness, instruction following, and fluency~\cite{li_2023_inference, wu_2025_axbench}.
Effectiveness varies across models and tasks~\cite{silva_2025_steering}, and interventions can shift individual inputs away from the intended response~\cite{tan_2024_analysing}.
The linear representation hypothesis provides a basis for studying these directions.
Concept-related variation can be expressed along directions in representation space~\cite{park_2024_linear, jiang_2024_origins}. 
Hierarchical child features can retain a shared parent component while adding orthogonal refinements~\cite{park_2025_geometry}.
Separable concepts are orthogonal under a causal inner product~\cite{park_2024_linear}.
These results motivate examining common components alongside category-specific repair variation.


\paragraph{Steering for Code Generation.} \label{sec:steering_for_code_generation}

Activation steering has been used to control programming language and library choices~\cite{rahman_2026_steering} and improve test diversity~\cite{zhu_2026_exploring}.
For security, SCS-Code builds directions from answer-position activations in security judgments~\cite{wendlinger_2026_security}.
Surgical Repair constructs directions from paired secure/insecure implementation instructions~\cite{sandoval_2026_surgical}.
MoC constructs directions from secure and vulnerable program pairs~\cite{yu_2025_mixture}.
DuoSteer combines separate safety and correctness directions at selected attention heads~\cite{yan_2026_interpreting}, while \textsc{Thea} identifies vulnerability-related sparse-autoencoder features~\cite{husseini_2026_repairing}.
However, adverse steering of security steering remain and cross-vulnerability transferability remains unclear~\cite{yan_2026_interpreting}. 
These studies motivate examining shared changes across repairs and their effects on generation.

\section{Conclusion} \label{sec:conclusion}

This study connects the representation of cryptographic API misuse repairs to safer code generation.
Matched program pairs reveal category-specific repair vectors with shared structure across languages and misuse categories. 
\method{} turns these shared changes into one steering direction and improves the proportion of outputs that are both secure and structurally valid across all five models. 
Refitting after source exclusion connects this control to repairs beyond the target language and misuse category. 
These findings establish controlled program repair as a useful basis for understanding and improving security-sensitive decisions in code models, and motivate shared repair directions for other vulnerabilities and complementary interventions for security and validity.

\section*{Limitations} \label{sec:limitations}

Our resource covers all ten cryptographic API misuse categories identified by~\citet{Mousavi_2024_investigation} in LLM-generated programs and four widely used programming languages~\cite{charisma_2026_top}, with four structural variants per original program. 
It is built from LLM-generated candidates and controlled rewrites. 
We study five 7B--16B instruction-tuned models with dense and Mixture-of-Experts (MoE) architectures. 
However, human-written programs, larger repositories, alternative implementations, and larger models remain to be evaluated.

The held-out alignment analysis uses the layer selected from the full corpus.
The source-exclusion steering also retains the full-source intervention scale.
Further independent evaluation would select layers and calibrate scales using only fitting data. 
Projection diagnostics use specific templates and API implementations.
Testing additional forms would assess their generality. 

Evaluation measures structural validity and compliance with the target security policy adapted from~\citet{Mousavi_2024_investigation}. 
But deployment requires broader security checks. 
Evaluator errors may affect results.
One author conducted an audit of the evaluators, but an independent second annotator would allow us to further assess evaluator reliability. 

In the comparison with other security steering methods, we adapt SCS-Code~\cite{wendlinger_2026_security} and Surgical Repair~\cite{sandoval_2026_surgical} with our dataset and contrasting instructions.
We use a common layer, generation settings, and $+1\sigma$ intervention strength for fair comparison of the steering vectors. 
However, each method's layer and intervention strength could be optimized separately to get a better result.

\section*{Ethical Considerations} \label{sec:ethical_considerations}
Our method aligns with the goal of making AI systems more helpful and harmless. 
By steering the outputs of code models, we contribute to reducing cryptographic API misuse in generated programs, thereby enhancing the safety and reliability of AI code assistants. 
However, reversing the activation steering direction lowers the security rate, which shows a dual-use risk.
The insecure program examples and negative interventions allow us to study models' internal representations and evaluate the effect of activation steering, but they could also be used to weaken code assistants. 
We encourage users of this technique to be responsible and avoid increasing unsafe or harmful behaviors via steering. 



\bibliography{custom}

\appendix

\section{Resource Construction}
\label{app:resource_construction}

\subsection{Dataset Organization and Task Coverage} \label{app:dataset_organization_and_task_coverage}
The resource is organized by the ten misuse categories M01--M10 in Table~\ref{tab:repair-categories}.
We allocate 300 originals per language--misuse category cell, giving every cell equal representation in direction construction. 
This design yields 1,200 originals per misuse category across the four languages and 12,000 originals in total. 
Programming tasks provide contexts in which these security decisions vary. 
The eight functionalities in Table~\ref{tab:app-task-catalogue} each provide three tasks, giving 24 tasks per language and 96 language-specific task texts. 
We distribute each cell's 300 originals evenly over its applicable tasks.
M01 covers symmetric encryption (F1) and message authentication codes (F6), while M04 covers asymmetric encryption (F3) and digital signatures (F4). 
These two misuse categories use six tasks per language, with 50 originals per task.
Other misuse categories use three tasks with 100 originals per task. 
When a task is relevant to multiple misuse categories, we allocate separate originals and target rewrites to each misuse category.
Therefore, all misuse categories receive the same total number of originals. 

\begin{table*}[t]
\centering
\footnotesize
\setlength{\tabcolsep}{1pt}
\begin{tabularx}{\textwidth}{@{}>{\raggedright\arraybackslash}p{0.37\textwidth}>{\raggedright\arraybackslash}Xl@{}}
\toprule
Functionality & Description & Categories \\
\midrule
F1: Symmetric encryption & Encrypt and decrypt data with the same secret key. & M01, M02 \\
F2: Symmetric encryption in CBC mode & Encrypt data using cipher block chaining (CBC) and an IV. & M03 \\
F3: Asymmetric encryption & Encrypt with a public key and decrypt with the private key. & M04 \\
F4: Digital signature & Produce a signature with a private key and verify it with a public key. & M04 \\
F5: Hash functions & Compute a fixed-length digest for detecting changes to data. & M09 \\
F6: Message authentication code (MAC) & Compute a keyed authentication tag to check integrity and authenticity. & M01, M10 \\
F7: Key derivation function (KDF) & Derive cryptographic key material from a password. & M06--M08 \\
F9: Pseudorandom number generator (PRNG) & Generate pseudorandom values for cryptographic operations. & M05 \\
\bottomrule
\end{tabularx}
\caption{Cryptographic functionalities from~\citet{Mousavi_2024_investigation}, retaining their IDs. Categories identify the targeted misuses in our resource.}
\label{tab:app-task-catalogue}
\end{table*}
\begin{table}[t]
\centering
\footnotesize
\setlength{\tabcolsep}{3pt}
\begin{tabular}{@{}lll@{}}
\toprule
Structure & Value placement & Construction \\
\midrule
\texttt{variable\_inline} & Variable & Inline \\
\texttt{variable\_helper} & Variable & Helper return \\
\texttt{callsite\_inline} & Call site & Inline \\
\texttt{callsite\_helper} & Call site & Helper return \\
\bottomrule
\end{tabular}
\caption{Four forms of each secure/misuse pair.}
\label{tab:app-rewrite-structures}
\end{table}

We retain the functionality and misuse IDs used by~\citet{Mousavi_2024_investigation} and adapt their Java tasks to the APIs of four languages.
Java uses standard cryptography APIs, TypeScript uses Node.js built-ins, C\# uses .NET APIs, and Python uses PyCA \texttt{cryptography} and standard-library cryptographic and randomness APIs.
The Java workflow for F8 (Key storage) functionality in~\citet{Mousavi_2024_investigation} has no direct counterpart shared by the native API families selected for all four languages. 
Therefore, we omit F8 functionality. 
Its target misuse M08 (hardcoded constant password) remains covered by F7 functionality. 
Their original M05 misuse concerns constant or predictable PRNG seeds.
However, caller-controlled seeding is not uniformly available in the cryptographic random APIs across the four languages. 
Therefore, we adapt M05 to the choice between a cryptographically secure random generator and a general-purpose PRNG for a security-sensitive value. 
Other misuse categories and classification policies follow~\citet{Mousavi_2024_investigation}.



\subsection{Generation, Rewriting, and Validation} \label{app:generation_rewriting_and_validation}
Original programs provide varied implementations from which to construct matched program pairs.
They are generated independently by five GPT models (\nolinkurl{gpt-4.1-mini-2025-04-14},
\nolinkurl{gpt-5.1-2025-11-13},
\nolinkurl{gpt-5.2-2025-12-11},
\nolinkurl{gpt-5.4-mini-2026-03-17}, and
\nolinkurl{gpt-5.6-luna}), with equal weights for program diversity. 
We remove duplicates and select originals that use the required APIs, implement the requested operation, and expose an editable target decision.

We provide each original and its rewrite rules to \nolinkurl{gpt-5.6-sol}.
The rewrite rules specify secure and misuse implementations while preserving the task, input/output interface, entry point, and surrounding operations.
Each original produces one secure and one misuse implementation in each of four structural forms.
The four structural forms cross two choices about the target value. 
As shown in Table~\ref{tab:app-rewrite-structures}, the target value is either assigned to a variable or supplied at a call site, and is either constructed inline or returned by a helper function.
Figure~\ref{fig:app-key-structures} gives an example: a key can be stored in a named variable before a cipher call, or supplied directly to that call.
Either form can construct the key locally or inside a helper function.
Thus, the structure change preserves the controlled security decision while varying how the value reaches the API. 
Each original contributes four matched secure/misuse pairs, or eight concrete programs.
Their representation stability is examined in Appendix~\ref{app:representation_stability_across_program_structures}.

\begin{figure*}[t]
\centering
\begin{minipage}[t][20mm][t]{0.48\textwidth}
\begin{lstlisting}[style=appendixprompt,language=Python]
key = secrets.token_bytes(32)
cipher = AESGCM(key)
\end{lstlisting}
\vfill
\centering\footnotesize (a) \texttt{variable\_inline}
\end{minipage}\hfill
\begin{minipage}[t][20mm][t]{0.48\textwidth}
\begin{lstlisting}[style=appendixprompt,language=Python]
cipher = AESGCM(secrets.token_bytes(32))
\end{lstlisting}
\vfill
\centering\footnotesize (b) \texttt{callsite\_inline}
\end{minipage}

\par\medskip
\begin{minipage}[t][30mm][t]{0.48\textwidth}
\begin{lstlisting}[style=appendixprompt,language=Python]
def create_key_material():
    return secrets.token_bytes(32)

key = create_key_material()
cipher = AESGCM(key)
\end{lstlisting}
\vfill
\centering\footnotesize (c) \texttt{variable\_helper}
\end{minipage}\hfill
\begin{minipage}[t][30mm][t]{0.48\textwidth}
\begin{lstlisting}[style=appendixprompt,language=Python]
def create_key_material():
    return secrets.token_bytes(32)

cipher = AESGCM(create_key_material())
\end{lstlisting}
\vfill
\centering\footnotesize (d) \texttt{callsite\_helper}
\end{minipage}
\caption{Example Python snippets showing the four structures for the same M01 key choice. Each secure form creates a 32-byte random key and constructs an AES-GCM cipher. The corresponding misuse form replaces \texttt{secrets.token\_bytes(32)} with  \texttt{'correct-horse-battery-staple-123'.encode('utf-8')}, keeping the structure unchanged. 
The package import, encryption and decryption operations are omitted in the example.}
\label{fig:app-key-structures}
\end{figure*}

We checked the quality of the corpus before including them in the representation corpus. 
We check parsing or compilation acceptance, imports, API use, required operations, API arguments, and operation sequence of each program in the dataset, as detailed in Appendix~\ref{app:structural_validity}. 
All 96,000 programs in the corpus are structurally valid. 
We also check all ten misuse categories and confirm that each misuse rewrite contains only the target misuse, and each secure rewrite contains the secure use of the corresponding API. 
For non-target categories, relevant API use is either secure or absent.
The first author manually inspected 3\% of programs in the dataset and confirmed that they were structurally valid and their API use matched their intended security label.

\subsection{Diagnostic Programs with Combined Security Choices}
\label{app:diagnostic_programs_with_combined_security_choices}
To examine how combining several API choices within a program affects its representation (Section~\ref{sec:response_to_individual_and_combined_security_choices}), we construct a separate diagnostic corpus. 
Each language uses one program template containing ten cryptographic operations, one for each misuse category. 
Each operation decision can be changed while keeping the other nine choices fixed. 
Enumerating all $2^{10}$ choices yields 1,024 programs per language and 4,096 programs overall, including all-secure and all-misuse endpoints. 
Appendix~\ref{app:controlled_program_diagnostics} describes the diagnostic analysis.

\section{Representation and Intervention Details}
\label{app:representation_and_intervention_details}

\subsection{Models and Program Representations}
\label{app:models_and_program_representations}
We use the released checkpoints and tokenizers in Section~\ref{sec:models_and_layer_selection} through Hugging Face Transformers~\cite{wolf_2020_transformers}.
Representation extraction uses bfloat16 weights and a batch size of one. 
Layer indices count decoder blocks from zero, excluding the embedding layer.
All selected layers are intermediate layers that contain the residual output after attention and feed-forward updates.

Each program is supplied as source code without a chat template in one forward pass. 
Let $h_{\ell,t}(x)\in\mathbb R^D$ denote the token activation at position $t$ and layer $\ell$ for program $x$, with hidden dimension $D$. 
We obtain its program representation $\bar h_\ell(x)$ by averaging token activations over the program body:
\begin{equation}
 \bar h_\ell(x)=\frac{1}{|I_{\mathrm{body}}(x)|}
           \sum_{t\in I_{\mathrm{body}}(x)}h_{\ell,t}(x),
 \label{eq:program-representation-readout}
\end{equation}
where $I_{\mathrm{body}}(x)$ contains the token positions overlapping the source code after the leading Java package/import declarations, TypeScript imports, C\# \texttt{using} directives, or Python imports.
The leading declarations remain in the input and provide context for later activations. 
Averaging the body positions gives every program a representation of the same dimension $D$ despite differences in source length.

\begin{table}[t]
\centering
\footnotesize
\setlength{\tabcolsep}{1pt}
\begin{tabular}{@{}lcccccS[table-format=2.2,table-number-alignment=center]@{}}
\toprule
Model & \shortstack{Para-\\meters} & \shortstack{Total\\layers} &
\shortstack{Selected\\layer} & \shortstack{AUROC\\ $J_\ell$} &
\shortstack{Selected\\PC} & {$\sigma$} \\
\midrule
\href{https://huggingface.co/meta-llama/Llama-3.1-8B-Instruct}{Llama}
& 8B & 32 & 14 & 0.783 & 1 & 1.45 \\
\href{https://huggingface.co/google/codegemma-7b-it}{CodeGemma}
& 7B & 28 & 18 & 0.803 & 1 & 3.74 \\
\href{https://huggingface.co/Qwen/Qwen2.5-7B-Instruct}{Qwen}
& 7B & 28 & 17 & 0.793 & 2 & 12.26 \\
\href{https://huggingface.co/Qwen/Qwen2.5-Coder-7B-Instruct}{Qwen-Coder}
& 7B & 28 & 16 & 0.821 & 1 & 17.58 \\
\href{https://huggingface.co/deepseek-ai/DeepSeek-Coder-V2-Lite-Instruct}{DeepSeek}
& 16B & 27 & 11 & 0.784 & 1 & 2.72 \\
\bottomrule
\end{tabular}
\caption{Model settings: layers indexed from zero, layer selection AUROC $J_\ell$ (Equation~\ref{eq:layer-selection}), PCs indexed from one, and intervention scales $\sigma$.}
\label{tab:app-model-settings}
\end{table}

\subsection{Linear Probes and Layer Selection}
\label{app:linear_probes_and_layer_selection}
We select a common layer by measuring how well linear classifiers trained on some misuse categories or languages predict security labels in the others~\cite{belinkov_2022_probing}.
At each layer, ten \emph{leave-one-category-out} folds train on nine misuse categories and evaluate on the remaining one, retaining all languages. 
Each fold contains 86,400 training and 9,600 evaluation program representations.
For four \emph{leave-one-language-out} folds, we train on three languages and evaluate on the remaining language, retaining all misuse categories. 
Each fold contains 72,000 training and 24,000 evaluation program representations.

The probes are $L_2$-regularized logistic regression classifiers.
For training representation $\bar h_j$, the predicted probability of the secure label is
\begin{equation}
 p_j=\frac{1}{1+\exp[-(w^\top\bar h_j+b)]},
 \label{eq:app-probe-probability}
\end{equation}
where $w$ and $b$ are the probe weights and intercept. 
With $y_j=1$ for secure and $y_j=0$ for misuse, the loss over $n$ training representations is
\begin{equation}
 \begin{aligned}
  \ell_j&=-y_j\log p_j-(1-y_j)\log(1-p_j),\\
  \mathcal L_C(w,b)&=\frac1n\sum_{j=1}^{n}\ell_j
                    +\frac{\lVert w\rVert_2^2}{2Cn}.
 \end{aligned}
 \label{eq:app-linear-probe-loss}
\end{equation}
This combines a binary cross-entropy loss with inverse regularization strength $C=1$.
We use the L-BFGS algorithm and optimization tolerance $10^{-4}$, and all fits converged within the 10,000-iteration limit.

We score each fold by the AUROC, which measures how well the probe ranks secure programs above misuse programs~\cite{tom_2006_introduction}.
The layer score gives equal total weight to misuse category and language transfer:
\begin{equation}
 J_\ell=\tfrac12\left(
 \tfrac1{10}\sum_{g=1}^{10}A^{\mathrm{misuse}}_{\ell,g}
 +\tfrac14\sum_{a=1}^{4}A^{\mathrm{language}}_{\ell,a}
 \right),
 \label{eq:layer-selection}
\end{equation}
where $A^{\mathrm{misuse}}_{\ell,g}$ and $A^{\mathrm{language}}_{\ell,a}$ are the AUROCs for held-out category $g$ and language $a$, respectively. 
We select the layer with the highest $J_\ell$.
Table~\ref{tab:app-model-settings} reports the selected layer and its $J_\ell$. 

\subsection{Direction Construction} \label{app:direction_construction}
For original-level repair vectors $d_i$ from Equation~\ref{eq:original-repair}, define $\operatorname{unit}(x)=x/\lVert x\rVert_2$, and $u_i=\operatorname{unit}(d_i)$.
With $N=12,000$ and $\mathcal I_c$ containing the 300 originals of cell $c$, the steering vectors for Cell DiffMean $v_{\mathrm{cell},c}$, Pooled DiffMean $v_{\mathrm{pooled}}$, and Normalized DiffMean $v_{\mathrm{norm}}$ are
\begin{equation}
\begin{aligned}
 v_{\mathrm{cell},c}&=\operatorname{unit}\!\left(
       \frac{1}{|\mathcal I_c|}\sum_{i\in\mathcal I_c}d_i\right),\\
 v_{\mathrm{pooled}}&=\operatorname{unit}\!\left(
       \frac1N\sum_{i=1}^{N}d_i\right),\\
 v_{\mathrm{norm}}&=\operatorname{unit}\!\left(
       \frac1N\sum_{i=1}^{N}u_i\right).
\end{aligned}
\label{eq:repair-normalized-mean}
\end{equation}
For \method{}, Equation~\ref{eq:repair-} defines the SVD objective. 
Let $\mathcal W=\{v_1,-v_1,\ldots,v_{10},-v_{10}\}$ be the set of twenty signed candidates with $w\in\mathcal W$, and let $\mathcal I_g$ index the repairs in misuse category $g$. 
We measure each candidate's alignment with repairs in every misuse category.
We define
\begin{equation}
\begin{aligned}
 a_g(w)&=\frac{1}{|\mathcal I_g|}
          \sum_{i\in\mathcal I_g}u_i^\top w,\\
 v_{\mathrm{CrypSteer}}&\in\arg\max_{w\in\mathcal W}
          \sum_{g=1}^{10}\mathbf{1}[a_g(w)>0],
\end{aligned}
\label{eq:repair-selection}
\end{equation}
where $\mathbf{1}[\cdot]$ is the indicator function, and $a_g(w)$ is candidate $w$'s mean cosine with category $g$'s repairs. 
A positive mean indicates alignment with the category's secure-minus-misuse changes. 
Here $v_{\mathrm{CrypSteer}}$ is the selected unit direction for \method{} steering.  
If multiple signed components rank for the first place, we select the component with the largest worst-category mean cosine.
Each vector has unit length, which allows for fair comparison at the same intervention strength.

\subsection{Scale and Generation Settings}
\label{app:scale_and_generation_settings}
We calibrate intervention strength from the spread of program representations within each model. 
For $N_{\mathrm{prog}}=96,000$ program representations $\bar h_j$ at the selected layer, let $\mu_h=\frac{1}{N_{\mathrm{prog}}}\sum_{j=1}^{N_{\mathrm{prog}}}\bar h_j$.
We use
\begin{equation}
 \sigma=\sqrt{\frac1{N_{\mathrm{prog}}-1}
              \sum_{j=1}^{N_{\mathrm{prog}}}
              \lVert\bar h_j-\mu_h\rVert_2^2}.
 \label{eq:generation-scale}
\end{equation}
Equivalently, $\sigma=\sqrt{\operatorname{tr}(\widehat\Sigma)}$, where $\widehat\Sigma$ is the sample covariance matrix~\cite{jolliffe_2016_principal}.

In the activation steering experiment, every unit direction receives the same perturbation norm $|\alpha|\sigma$ with a fixed $\alpha$. 
The activation steering hook adds $\alpha\sigma v$ to token activations at the selected decoder block's output and updates the last prompt token and every decoding position. 
Each condition generates 30 attempts for each of five prompts per cell.
Generation uses the model's chat template with one user message, temperature 0.8, top-$p$ 0.95, and at most 2,048 new tokens. 
Seeds 2000--2029 are reused across conditions for the 30 attempts.

\begin{figure}[t]
\centering
\includegraphics[width=\columnwidth]{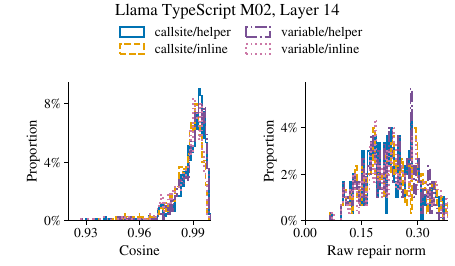}
\par\nointerlineskip\vspace{1pt}
\begin{minipage}[t]{0.49\columnwidth}
  \centering
  {\footnotesize \qquad \qquad (a) Direction cosine\par}
\end{minipage}\hfill
\begin{minipage}[t]{0.49\columnwidth}
  \centering
  {\footnotesize \qquad \qquad(b) Repair norm\par}
\end{minipage}
\caption{Structure agreement for 300 originals in Llama layer 14 TypeScript--M02 cell. (a) Each structure contributes 900 cosine values from comparisons with the other three structures within each original. (b) Each structure contributes 300 raw repair norms.}
\label{fig:app-structure-distributions}
\end{figure}

\section{Representation Stability Across Program Structures}
\label{app:representation_stability_across_program_structures}
In the dataset, the controlled rewrites for each original appear in four code structures, which we average before fitting a shared direction. 
This provides a controlled way to examine how program organization affects its repair vector.
We examine structure-specific repair vectors  $\delta_{i,s}$ to assess their stability to support our choice of averaging repair vectors for four program structures before fitting a shared direction.


\subsection{An Illustration at the Original Level} \label{app:an_illustration_at_the_original_level}
We first inspect whether the four repair vectors for each original point in similar directions and have similar lengths.
We give an example in Figure~\ref{fig:app-structure-distributions} to show what agreement looks like. 
The example is the TypeScript-M02 cell at Llama's selected layer. 
For each of the 300 originals, four vectors yield six distinct pairwise cosine similarities. 
Figure~\ref{fig:app-structure-distributions}(a) groups these comparisons by structure. 
Each distribution compares one structure with the other three across 300 originals, giving $300\times3=900$ cosine values.
Each pairwise comparison appears in two curves, one for each structure. 
Across the $300\times6=1,800$ comparisons, the mean cosine is 0.988, indicating close alignment. 

Figure~\ref{fig:app-structure-distributions}(b) shows the repair vector norms, with 300 values per structure. 
These distributions include differences in repair length between originals.
To isolate variation among the four forms of the same original, we compute their coefficient of variation (CV), which is the population standard deviation divided by their mean. 
Writing $r_{i,s}=\lVert\delta_{i,s}\rVert_2$ and $\bar r_i=\frac14\sum_s r_{i,s}$, the CV is
\begin{equation}
 \mathrm{CV}_i=
 \frac{\sqrt{\frac14\sum_{s=1}^4(r_{i,s}-\bar r_i)^2}}{\bar r_i}.
 \label{app:eq-structure-norm-cv}
\end{equation}
The average CV across the 300 originals is 1.36\%.
Therefore, within a cell, changing code structure has little effect on either the direction or the length of the repair vector, which supports averaging the four vectors to represent the repair of each original program.

\subsection{Category Correspondence Across Structures} \label{app:category_correspondence_across_structures}
We next test whether category-level directions remain recognizable across structures throughout the resource.
Given a misuse category's direction in one structure, we find its closest match among ten category references built from the other three structures in the same language. 
For example, the TypeScript M02 direction should be closest to the M02 reference among all misuse category references. 
For language $l$, misuse category $g$, and structure $s$, the direction to be matched is
\begin{equation}
 a_{l,g,s}=\operatorname{unit}\!\left(
       \frac1{300}\sum_{i\in(l,g)}\delta_{i,s}\right).
 \label{app:eq-structure-query}
\end{equation}
where $\operatorname{unit}(x)=x/\lVert x\rVert_2$. 
We treat $a_{l,g,s}$ as the \emph{query}.
We hold out structure $s$, and construct ten \emph{references}, one for each candidate misuse category $k$, by averaging its unit directions from the other three structures in the same language. 
\begin{equation}
 c_{l,k,-s}=\operatorname{unit}\!\left(
                \frac13\sum_{r\ne s}a_{l,k,r}\right).
 \label{app:eq-structure-reference}
\end{equation}
We assign the category with the highest cosine similarity $\widehat g=\arg\max_{k\in\{1,\ldots,10\}} a_{l,g,s}^{\top}c_{l,k,-s}$.
This dot product is the cosine similarity as they have unit length.
Retrieval is correct when $\widehat g=g$.
Repeating this procedure for four languages, ten misuse category queries, and four structures gives  $4\times10\times4=160$ queries per model. 
The results show that all models correctly recover the category for 159 or 160 queries.
Therefore, the identity of the misuse category is largely preserved in the four structures. 
This supports averaging the four repair vectors within each original to summarize their signals before constructing a shared direction.

\section{Additional Geometry Analysis and Clustering}
\label{app:additional_geometry_analysis_and_clustering}
We conduct additional geometry analysis to examine whether repair vectors are organized by misuse category across languages, how much language changes that organization, and whether it persists across layers. 
For each cell with language $l$ and misuse category $g$, let $\mathcal I_{l,g}$ index its 300 originals.
We average repair differences $d_i$ from Equation~\ref{eq:original-repair} to get the cell's mean repair difference $\bar d_{l,g}$, then normalize the resulting mean to get its unit-length direction $v_{\mathrm{cell},l,g}$, which is the Cell DiffMean direction $v_{\mathrm{cell},c}$ in Appendix~\ref{app:direction_construction} with $c=(l,g)$ written explicitly.
\begin{equation}
 \begin{aligned}
 \bar d_{l,g}&=\frac1{300}\sum_{i\in\mathcal I_{l,g}}d_i,\\
 v_{\mathrm{cell},l,g}&=\frac{\bar d_{l,g}}{\lVert\bar d_{l,g}\rVert_2}.
 \end{aligned}
 \label{app:eq-cell-mean-direction}
\end{equation}
Normalization preserves the orientation while removing its magnitude, which allows comparison of the 40 cell directions independently of their lengths.

\begin{figure}[t]
\centering
\includegraphics[width=0.9\columnwidth]{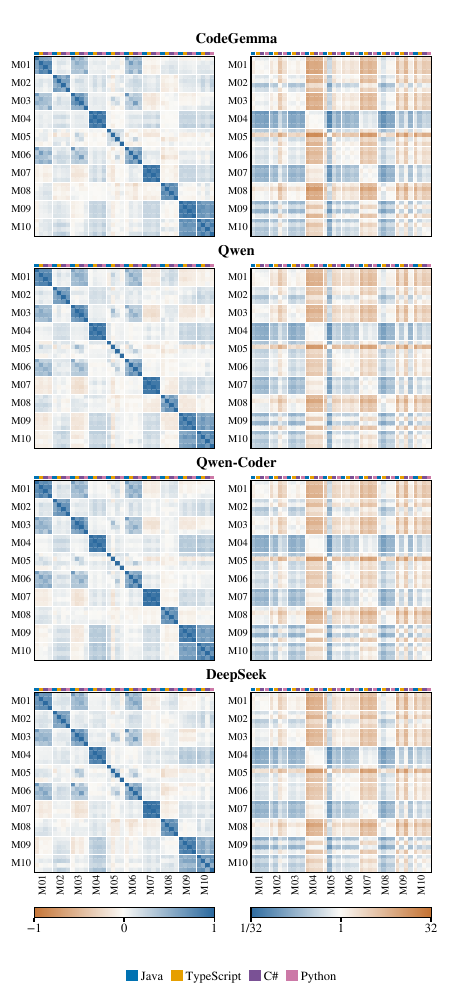}
\par\nointerlineskip\vspace{1pt}
\begin{minipage}[t]{0.45\columnwidth}
  \centering
  {\footnotesize \qquad (a) Direction cosine\par}
\end{minipage}\hfill
\begin{minipage}[t]{0.45\columnwidth}
  \centering
  {\footnotesize \qquad (b) Norm ratio\par}
\end{minipage}
\caption{Repair representation geometry for the other four models. Panel definitions and color scales follow Figure~\ref{fig:repair-geometry}.}
\label{app:fig-cell-cosines}
\label{app:fig-cell-norms}
\end{figure}

\subsection{Category Alignment and Repair Magnitude} \label{app:category_alignment_and_repair_magnitude}
We first use cosine similarity to test whether the same misuse category repair has a recognizable direction across different languages. 
We compare two sets of pairs.
First, the same misuse category in different languages, which gives $10\times\binom42=60$ pairs.
Second, different misuse categories within one language, which gives $4\times\binom{10}2=180$ pairs.
The two mean-cosine columns in Table~\ref{table-geometry-summary} report the mean cosine for same category different language and same language different category pairs. 
The high same category mean cosine values indicate that the misuse category correspondence extends across programming languages.

We next use a retrieval test to ask whether a cell's category can be identified from directions in other languages.
For language $l$ and misuse category $g$, the query is $v_{\mathrm{cell},l,g}$.
We build ten references using the other three languages and choose the one that has the highest cosine similarity to it:
\begin{equation}
 \begin{aligned}
 v_{\mathrm{ref},-l,k}&=\operatorname{unit}\!\left(
                  \frac13\sum_{b\ne l}v_{\mathrm{cell},b,k}\right),\\
 \widehat g&=\arg\max_{k\in\{1,\ldots,10\}}
                   v_{\mathrm{cell},l,g}^{\top}v_{\mathrm{ref},-l,k}.
 \end{aligned}
 \label{app:eq-language-category-retrieval}
\end{equation}
where $\operatorname{unit}(x)=x/\lVert x\rVert_2$, and $v_{\mathrm{ref},-l,k}$ is the unit reference direction for misuse category $k$, formed without language $l$.
A match is correct when $\widehat g=g$.
There are 40 queries per model, one for each language-misuse category cell. 
The results are shown in the correct category matches column in Table~\ref{table-geometry-summary}.
All six incorrect matches concern M05, which could be explained by the adaptation of M05 definition from~\citet{Mousavi_2024_investigation} and M05 concerns about choosing to use cryptographic API instead of misuse a cryptographic API.

We also analyze the repair length because direction and length answer different questions. 
We average each original's repair representation length within each cell to get a scalar mean $\rho_{l,g}$ of them:
\begin{equation}
 \rho_{l,g}=\frac1{300}\sum_{i\in\mathcal I_{l,g}}\lVert d_i\rVert_2.
 \label{app:eq-mean-repair-length}
\end{equation}
The last column in Table~\ref{table-geometry-summary} shows the $\max_{l,g}\rho_{l,g}/\min_{l,g}\rho_{l,g}$, which compares the largest and smallest cell-average repair representation lengths. 
The substantial ratios show why we need to use unit repairs to construct a shared direction because the raw repair representations gives longer repairs more influence.

In Section~\ref{sec:representation_organization_Across_languages_and_misuse_categories}, we show repair representation geometry heatmaps for the selected layer for Llama. 
Figure~\ref{app:fig-cell-cosines} extends Figure~\ref{fig:repair-geometry} to the other four models. 
Figure~\ref{app:fig-cell-cosines}(a) shows cross-language alignment by strong off-diagonal entries inside the diagonal blocks corresponding to individual misuse categories. 
Alignment also appears between some categories, such as M09 and M10. 
The horizontal and vertical bands in the norm ratio heatmap in Figure~\ref{app:fig-cell-norms}(b) show scale differences for different misuse categories.

\subsection{Category and Language Main Effects} \label{app:category_and_language_main_effects}

We use the sum-of-squares decomposition of balanced two-way analysis of variance (ANOVA) to quantify the variation associated with misuse category and programming language~\cite{andrew_2005_analysis}.
This analysis examines whether repair vectors are organized more strongly by misuse category or by programming language. 
Write $v_{\mathrm{cell},l,g}=v_{\mathrm{cell},c}$ for cell $c=(l,g)$.
Let $\bar v$ denote the mean of the 40 unit cell repair vectors, $\bar v_{l,\cdot}$ average the ten misuse category repair vectors in language $l$, and $\bar v_{\cdot,g}$ average misuse category $g$ across four languages. 
We compare the total centered variation $T$, the variation between misuse category means $SS_G$, and the variation between language means $SS_L$:
\begin{equation}
 \begin{aligned}
 T&=\sum_{l,g}\lVert v_{\mathrm{cell},l,g}-\bar v\rVert_2^2,\\
 SS_G&=4\sum_g\lVert\bar v_{\cdot,g}-\bar v\rVert_2^2,\\
 SS_L&=10\sum_l\lVert\bar v_{l,\cdot}-\bar v\rVert_2^2.
 \end{aligned}
 \label{app:eq-geometry-factor-effects}
\end{equation}
The category term $SS_G$ measures separation between misuse category means after averaging over languages. 
The language term $SS_L$ measures separation between language means after averaging over misuse categories. 
The factors 4 and 10 account for the number of cells represented by each mean. 
Table~\ref{app:tab-geometry-factors} reports the category and language sums of squares as percentages of the total variation $100SS_G/T$ and $100SS_L/T$, respectively.
The larger category effect supports the category organization observed in the cosine similarity analysis. 

\begin{table}[t]
\centering
\footnotesize
\setlength{\tabcolsep}{1pt}
\begin{tabular}{@{}lcc@{}}
\toprule
Model & Category main effect &
Language main effect\\
\midrule
Llama & 64.42\% & 5.52\%\\
CodeGemma & 70.80\% & 4.67\%\\
Qwen & 71.24\% & 4.67\%\\
Qwen-Coder & 72.39\% & 4.57\%\\
DeepSeek & 68.86\% & 5.17\%\\
\bottomrule
\end{tabular}
\caption{Category and language shares of total centered variation among 40 unit cell directions.}
\label{app:tab-geometry-factors}
\end{table}

\begin{table}[t]
\centering
\footnotesize
\setlength{\tabcolsep}{4pt}
\begin{tabular}{@{}lcc@{}}
\toprule
Model & Category ARI & Language ARI\\
\midrule
Llama & 0.565 & $-0.046$\\
CodeGemma & 0.657 & $-0.064$\\
Qwen & 0.771 & $-0.110$\\
Qwen-Coder & 0.798 & $-0.112$\\
DeepSeek & 0.895 & $-0.119$\\
\bottomrule
\end{tabular}
\caption{Agreement of the $k=10$ clustering of 40 unit cell directions with category and language labels at the selected layer.}
\label{app:tab-clustering}
\end{table}

\subsection{Clustering of Cell Directions} \label{app:clustering_of_cell_directions}
\begin{figure}[t]
\centering
\includegraphics[width=\columnwidth]{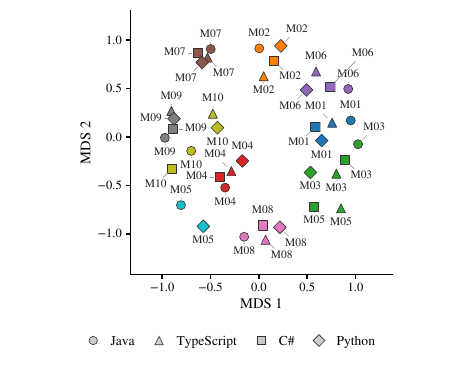}
\caption{DeepSeek cell directions at the selected layer, visualized with MDS. Colors denote clusters, shapes indicate languages, and point labels identify misuse categories.}
\label{app:fig-clustering-scatter}
\end{figure}

We group 40 unit cell repair vectors in their original representation space using Euclidean $k$-means with $k=10$. 
We measure correspondence with category and language labels using the adjusted Rand index (ARI)~\cite{hubert_1985_comparing}.
An ARI of 1 indicates that the clustering exactly matches the reference grouping, while scores around 0 indicate agreement comparable to random grouping. 
As shown in Table~\ref{app:tab-clustering}, the partitions correspond closely to misuse categories. 
Figure~\ref{app:fig-clustering-scatter} uses metric multidimensional scaling (MDS) to approximate DeepSeek's pairwise direction distances in two dimensions~\cite{mair_2022_more}.
The clusters largely bring together the four languages of the same category.

\begin{figure*}[t]
\centering
\includegraphics[width=\textwidth]{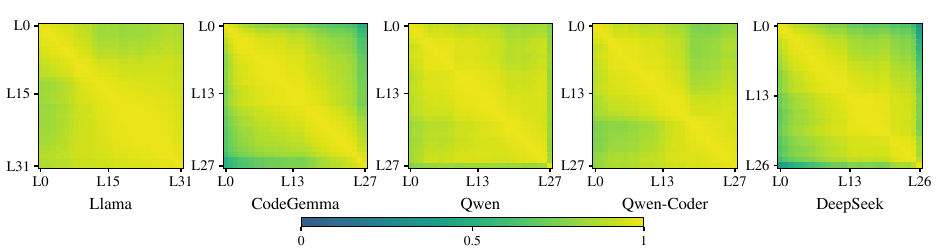}
\caption{Cross-layer repair geometry in all five models.}
\label{app:fig-layer-geometry}
\end{figure*}

\subsection{Geometry Stability Across Layers} \label{app:geometry_stability_across_layers}

Following representational similarity analysis~\cite{kriegeskorte_2008_representational}, we compare the cell-to-cell similarity across layers. 
For each layer, we collect the strict upper triangle of its $40\times40$ cell-direction cosine matrix into a vector.
The resulting $\binom{40}2=780$ entries describe the same cell pairs at every layer. 
Cosine similarity between two such vectors measures agreement between the two layers' relational geometry.
We use Figure~\ref{app:fig-layer-geometry} to show the layer-wise relational geometry. 
High similarity across layers suggests that the geometry of repair representations persists in different layers. 
Figure~\ref{app:fig-layer-geometry} shows the results, suggesting stable geometric relationships between repair vectors across most of the layers, particularly from early-middle to late layers.

\begin{figure*}[t]
\centering
\includegraphics[width=\textwidth]{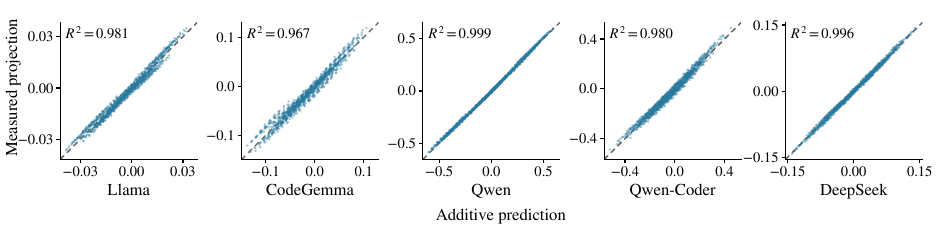}
\caption{Full-configuration additive fits for Java across all five models, using their selected layer and shared direction. Each panel contains 1,024 combinations. Dashed lines mark equality.}
\label{fig:repair-diagnostics}
\end{figure*}

\begin{table*}[t]
\centering
\scriptsize
\setlength{\tabcolsep}{3pt}
\begin{tabular}{@{}ll*{3}{r@{\hspace{0.25em}}l}@{}}
\toprule
Model & Comparator &
\multicolumn{2}{c}{$\Delta$\texttt{secure@1}} &
\multicolumn{2}{c}{$\Delta$\texttt{valid@1}} &
\multicolumn{2}{c}{$\Delta$\texttt{secure-valid@1}} \\
\midrule

\multirow{4}{*}{Llama} & Baseline
& $15.00$ & $[12.83,17.18]^*$
& $-3.17$ & $[-5.13,-1.25]^*$
& $9.67$ & $[7.35,11.98]^*$ \\
& Cell DiffMean
& $5.18$ & $[3.25,7.12]^*$
& $1.27$ & $[-0.68,3.22]$
& $4.85$ & $[2.75,6.95]^*$ \\
& Pooled DiffMean
& $4.97$ & $[3.33,6.62]^*$
& $-0.52$ & $[-2.28,1.22]$
& $2.78$ & $[0.87,4.68]^*$ \\
& Normalized DiffMean
& $2.20$ & $[0.95,3.47]^*$
& $0.03$ & $[-1.50,1.53]$
& $0.85$ & $[-0.73,2.43]$ \\
\midrule

\multirow{4}{*}{CodeGemma} & Baseline
& $10.42$ & $[8.28,12.60]^*$
& $1.38$ & $[-0.92,3.68]$
& $5.20$ & $[3.23,7.17]^*$ \\
& Cell DiffMean
& $2.62$ & $[0.73,4.47]^*$
& $-0.13$ & $[-2.27,2.02]$
& $0.87$ & $[-1.05,2.78]$ \\
& Pooled DiffMean
& $2.95$ & $[0.93,4.97]^*$
& $2.35$ & $[0.38,4.38]$
& $2.23$ & $[0.47,4.03]$ \\
& Normalized DiffMean
& $1.23$ & $[-0.77,3.28]$
& $0.93$ & $[-0.88,2.78]$
& $0.58$ & $[-1.12,2.30]$ \\
\midrule

\multirow{4}{*}{Qwen} & Baseline
& $10.80$ & $[8.58,13.08]^*$
& $-2.68$ & $[-4.83,-0.55]$
& $5.70$ & $[3.18,8.27]^*$ \\
& Cell DiffMean
& $3.32$ & $[1.47,5.18]^*$
& $1.23$ & $[-1.03,3.48]$
& $1.87$ & $[-0.33,4.10]$ \\
& Pooled DiffMean
& $2.87$ & $[1.60,4.13]^*$
& $-0.03$ & $[-1.43,1.40]$
& $1.88$ & $[0.50,3.32]^*$ \\
& Normalized DiffMean
& $-0.98$ & $[-2.12,0.12]$
& $-1.10$ & $[-2.55,0.37]$
& $-1.05$ & $[-2.37,0.28]$ \\
\midrule

\multirow{4}{*}{Qwen-Coder} & Baseline
& $5.28$ & $[3.75,6.82]^*$
& $-1.67$ & $[-3.47,0.15]$
& $2.32$ & $[0.55,4.10]$ \\
& Cell DiffMean
& $1.75$ & $[0.45,3.08]^*$
& $-0.53$ & $[-2.22,1.12]$
& $0.97$ & $[-0.70,2.67]$ \\
& Pooled DiffMean
& $1.02$ & $[-0.20,2.27]$
& $0.27$ & $[-1.37,1.90]$
& $1.50$ & $[-0.12,3.13]$ \\
& Normalized DiffMean
& $-0.18$ & $[-1.23,0.90]$
& $0.92$ & $[-0.42,2.27]$
& $0.62$ & $[-0.75,1.98]$ \\
\midrule

\multirow{4}{*}{DeepSeek} & Baseline
& $6.85$ & $[4.35,9.33]^*$
& $-2.20$ & $[-3.97,-0.45]$
& $5.03$ & $[2.40,7.67]^*$ \\
& Cell DiffMean
& $-0.32$ & $[-2.97,2.28]$
& $0.52$ & $[-1.25,2.27]$
& $0.07$ & $[-2.53,2.65]$ \\
& Pooled DiffMean
& $1.10$ & $[-0.18,2.42]$
& $0.00$ & $[-1.23,1.23]$
& $0.88$ & $[-0.53,2.35]$ \\
& Normalized DiffMean
& $0.25$ & $[-0.97,1.50]$
& $0.25$ & $[-0.88,1.37]$
& $0.25$ & $[-1.13,1.57]$ \\
\bottomrule
\end{tabular}
\caption{\method{} minus each comparator at $+1\sigma$.
The baseline is unsteered. Differences are in percentage points,
with 95\% confidence intervals over 6,000 paired attempts.
$^*$ denotes Holm-adjusted $p<0.05$.}
\label{tab:app-key-contrasts}
\end{table*}

\section{Controlled Program Diagnostics} \label{app:controlled_program_diagnostics}
\subsection{Additive Fits and Combination Prediction}
\label{app:additive_fits_and_combination_prediction}
The diagnostic programs in Appendix~\ref{app:diagnostic_programs_with_combined_security_choices} vary ten security choices within one template per language.
Let $z\in\{0,1\}^{10}$ encode the configuration, with $z_g=1$ for the secure implementation of category $g$ and $z_g=0$ for misuse.
We fix each model's unit direction $v_{\mathrm{CrypSteer}}$ for the analysis.
We center projections at the midpoint between the all-misuse and all-secure endpoints.
Writing $\mathbf0$ and $\mathbf1$ for these two configurations, and  $\bar h_{\mathrm{mid}}=(\bar h(\mathbf0)+\bar h(\mathbf1))/2$, we calculate the centered projection:
\begin{equation}
 p(z)=\left[\bar h(z)-\bar h_{\mathrm{mid}}\right]^\top
             v_{\mathrm{CrypSteer}}.
 \label{app:eq-composition-projection}
\end{equation} 
We use midpoint centering to make visualization intuitive while preserving projection differences.

\paragraph{Descriptive additive fit.}
We fit and evaluate the additive model $\widehat p_{\mathrm{fit}}(z)=\beta_0+\sum_{g=1}^{10}\beta_gz_g$ on all 1,024 configurations using ordinary least squares, with intercept $\beta_0$ and category contribution $\beta_g$ that minimize the sum of squared differences between fitted and measured projections.
Figure~\ref{fig:repair-diagnostics} shows the Java fits for all five models, and each point is one configuration. 
We can find that all points are close to the equality line, which indicates a close fit. 

\paragraph{Prediction from individual repairs.}
Next, using the all-misuse program and ten single-repair programs, we predict the projection of each remaining combination as 
\begin{equation}
 \widehat p_{\mathrm{pred}}(z)=p(\mathbf0)+\sum_{g=1}^{10}z_g
                       [p(e_g)-p(\mathbf0)],
 \label{app:eq-anchor-prediction}
\end{equation}
where $e_g$ is the configuration in which only operation $g$ is secure.
We evaluate these predictions on the remaining 1,013 configurations. 
Table~\ref{tab:combination-prediction} shows the $R^2$ for the prediction. 
The results show that responses to individual repairs can approximate those of combined repairs.

\subsection{Paired Code Heatmaps} \label{app:paired_code_heatmaps}
In Section~\ref{sec:response_to_individual_and_combined_security_choices}, we use a paired code heatmap to show how cryptographic API repairs change token projections onto the shared direction at both edited expressions and unchanged downstream uses.
When computing the token projection, the complete program enters the model, and we show excerpts. 
At the selected layer, for token $t$ in program $x$, its scalar projection is
\begin{equation}
 p_t(x)=h_{t}(x)^\top v_{\mathrm{CrypSteer}},
 \label{app:eq-token-projection}
\end{equation}
where $h_{t}(x)$ is the token activation and $v_{\mathrm{CrypSteer}}$ is the unit steering direction.

We align corresponding lines and lexical spans, such as identifiers and API expressions, across the paired programs.
The span alignment allows us to compare corresponding source expressions when their tokenization differs. 
For each aligned span pair $a$ and program version $y\in\{\mathrm{secure},\mathrm{misuse}\}$, the span projection is
\begin{equation}
 \bar p_a^y=\frac{\sum_t w_{a,t}^y\,p_t(x^y)}{\sum_t w_{a,t}^y},
 \label{app:eq-span-projection}
\end{equation}
where $x^y$ is the corresponding program and $w_{a,t}^y$ counts the source characters shared by token $t$ and the span.
For visualization, the paired difference is
\begin{equation}
 \Delta_a=\bar p_a^{\mathrm{secure}}-\bar p_a^{\mathrm{misuse}}.
 \label{app:eq-paired-projection}
\end{equation}
We color spans with $|\Delta_a|>0.2$, and display $\Delta_a$ on the secure side and $-\Delta_a$ on the misuse side, with colors clipped at $\pm0.5$.
Green denotes positive differences, and red denotes negative differences. 
A highlighted unchanged span identifies a context-dependent activation change at that location.

\begin{table*}[tp]
\centering
\scriptsize
\setlength{\tabcolsep}{0.5pt}
\begin{tabular*}{\textwidth}{@{\extracolsep{\fill}}ll*{5}{ccc}@{}}
\toprule
 & & \multicolumn{3}{c}{Baseline} & \multicolumn{3}{c}{\shortstack{Cell\\DiffMean}} & \multicolumn{3}{c}{\shortstack{Pooled\\DiffMean}} & \multicolumn{3}{c}{\shortstack{Normalized\\DiffMean}} & \multicolumn{3}{c}{\method{}} \\
\cmidrule(lr){3-5}\cmidrule(lr){6-8}\cmidrule(lr){9-11}\cmidrule(lr){12-14}\cmidrule(l){15-17}
Model & Language & \shortstack{\texttt{secure}\\\texttt{@1}} & \shortstack{\texttt{valid}\\\texttt{@1}} & \shortstack{\texttt{secure-}\\\texttt{valid@1}} & \shortstack{\texttt{secure}\\\texttt{@1}} & \shortstack{\texttt{valid}\\\texttt{@1}} & \shortstack{\texttt{secure-}\\\texttt{valid@1}} & \shortstack{\texttt{secure}\\\texttt{@1}} & \shortstack{\texttt{valid}\\\texttt{@1}} & \shortstack{\texttt{secure-}\\\texttt{valid@1}} & \shortstack{\texttt{secure}\\\texttt{@1}} & \shortstack{\texttt{valid}\\\texttt{@1}} & \shortstack{\texttt{secure-}\\\texttt{valid@1}} & \shortstack{\texttt{secure}\\\texttt{@1}} & \shortstack{\texttt{valid}\\\texttt{@1}} & \shortstack{\texttt{secure-}\\\texttt{valid@1}} \\
\midrule
Llama & Java & 53.53 & 59.87 & 33.47 & 74.33 & 55.47 & 42.60 & 76.73 & 59.00 & 46.53 & \underline{81.00} & 57.67 & \textbf{48.93} & \textbf{81.33} & 57.00 & \underline{48.33} \\
 & TypeScript & 62.00 & 73.47 & 48.47 & 68.07 & 66.60 & 52.87 & 63.80 & 67.53 & 52.73 & \underline{68.27} & 67.67 & \underline{55.20} & \textbf{72.07} & 67.47 & \textbf{56.47} \\
 & C\# & 62.67 & 70.47 & 43.93 & 70.67 & 68.93 & 49.47 & 71.80 & 67.33 & 49.40 & \underline{72.93} & 67.87 & \underline{52.20} & \textbf{74.47} & 68.07 & \textbf{52.27} \\
 & Python & 70.13 & 86.87 & 64.20 & 74.53 & 81.93 & 64.40 & 76.13 & 86.20 & 68.93 & \underline{77.33} & 84.67 & \underline{69.00} & \textbf{80.47} & 85.47 & \textbf{71.67} \\
\midrule
CodeGemma & Java & 56.20 & 37.87 & 25.80 & 68.40 & 39.07 & \underline{30.93} & 68.33 & 37.33 & 29.47 & \underline{70.67} & 37.40 & 30.47 & \textbf{72.47} & 39.93 & \textbf{32.00} \\
 & TypeScript & 49.67 & 49.87 & 28.40 & 56.47 & 52.07 & \underline{33.53} & \underline{58.73} & 48.53 & 33.07 & 57.93 & 48.87 & 33.13 & \textbf{62.20} & 51.20 & \textbf{36.27} \\
 & C\# & 47.27 & 23.20 & 13.67 & \underline{51.73} & 24.60 & \textbf{15.93} & 50.00 & 21.87 & 13.93 & 51.13 & 22.27 & \underline{14.87} & \textbf{52.07} & 19.80 & 14.40 \\
 & Python & 31.93 & 40.73 & 17.47 & 39.67 & 42.00 & 22.27 & 37.87 & 40.07 & 20.73 & \textbf{42.07} & 44.93 & \textbf{25.33} & \underline{40.00} & 46.27 & \underline{23.47} \\
\midrule
Qwen & Java & 53.60 & 75.60 & 42.93 & 64.00 & 67.60 & 46.73 & 66.00 & 72.33 & 49.53 & \textbf{70.73} & 72.87 & \underline{51.53} & \underline{68.53} & 73.53 & \textbf{52.13} \\
 & TypeScript & 56.87 & 65.67 & 40.33 & 59.13 & 58.13 & 38.20 & 60.47 & 59.07 & \underline{38.40} & \textbf{63.87} & 59.33 & \textbf{39.87} & \underline{62.47} & 57.47 & \underline{38.40} \\
 & C\# & 60.33 & 76.33 & 47.73 & 68.20 & 76.60 & \textbf{53.73} & 67.67 & 72.80 & 49.33 & \underline{69.20} & 73.87 & \underline{51.53} & \textbf{70.67} & 72.53 & 50.87 \\
 & Python & 53.87 & 81.13 & 48.33 & 63.27 & 80.73 & 56.00 & 62.27 & 83.93 & 57.33 & \textbf{68.00} & 86.33 & \textbf{63.40} & \underline{66.20} & 84.47 & \underline{60.73} \\
\midrule
Qwen-Coder & Java & 61.67 & 75.00 & 45.07 & 71.07 & 71.27 & \underline{49.33} & 70.20 & 64.27 & 44.73 & \underline{73.20} & 65.27 & 47.80 & \textbf{75.53} & 68.93 & \textbf{51.13} \\
 & TypeScript & 63.07 & 65.47 & 43.13 & 65.80 & 60.33 & 41.93 & \underline{66.47} & 61.07 & \underline{42.47} & \textbf{67.27} & 59.40 & 41.73 & 65.67 & 60.73 & \textbf{42.87} \\
 & C\# & 70.87 & 74.87 & 55.53 & 71.67 & 77.87 & 56.73 & 72.13 & 78.13 & \underline{57.27} & \underline{73.00} & 77.27 & \textbf{58.00} & \textbf{73.13} & 76.47 & 56.67 \\
 & Python & 61.33 & 79.93 & 58.60 & 62.53 & 81.27 & 59.73 & \underline{65.20} & 84.07 & \underline{61.13} & \textbf{65.33} & 83.00 & \textbf{61.60} & 63.73 & 82.47 & 60.93 \\
\midrule
DeepSeek & Java & 61.13 & 94.33 & 57.93 & \textbf{71.73} & 91.33 & \textbf{66.20} & 68.67 & 94.47 & 64.40 & 69.67 & 93.33 & 65.00 & \underline{70.27} & 93.40 & \underline{65.40} \\
 & TypeScript & 62.87 & 80.27 & 52.53 & \underline{68.20} & 76.67 & \textbf{54.60} & 66.53 & 72.07 & 51.47 & 67.60 & 73.13 & 52.33 & \textbf{68.40} & 73.87 & \underline{53.27} \\
 & C\# & 62.67 & 90.00 & 56.20 & 66.13 & 88.47 & 58.40 & 66.93 & 88.80 & 60.07 & \textbf{67.80} & 90.40 & \textbf{62.00} & \underline{67.13} & 89.80 & \underline{61.13} \\
 & Python & 60.80 & 91.87 & 56.67 & \textbf{70.07} & 89.13 & \textbf{64.00} & 68.33 & 92.33 & \textbf{64.00} & 68.80 & 89.80 & 63.13 & \underline{69.07} & 90.60 & \underline{63.67} \\
\bottomrule
\end{tabular*}
\caption{
Single-generation success rates over 1,500 outputs per model and condition for each language.
Bold and underlining mark the highest and second-highest \texttt{secure@1} and \texttt{secure-valid@1} among the four directions at $+1\sigma$.}
\label{tab:app-language-method-comparison}
\end{table*}

\begin{table*}[!t]
\centering
\scriptsize
\setlength{\tabcolsep}{0.1pt}
\begin{tabular*}{\textwidth}{@{\extracolsep{\fill}}ll*{5}{ccc}@{}}
\toprule
 & & \multicolumn{3}{c}{Baseline} & \multicolumn{3}{c}{Cell DiffMean} & \multicolumn{3}{c}{Pooled DiffMean} & \multicolumn{3}{c}{Normalized DiffMean} & \multicolumn{3}{c}{\method{}} \\
\cmidrule(lr){3-5}\cmidrule(lr){6-8}\cmidrule(lr){9-11}\cmidrule(lr){12-14}\cmidrule(l){15-17}
Model & Category & \shortstack{\texttt{secure}\\\texttt{@1}} & \shortstack{\texttt{valid}\\\texttt{@1}} & \shortstack{\texttt{secure-}\\\texttt{valid@1}} & \shortstack{\texttt{secure}\\\texttt{@1}} & \shortstack{\texttt{valid}\\\texttt{@1}} & \shortstack{\texttt{secure-}\\\texttt{valid@1}} & \shortstack{\texttt{secure}\\\texttt{@1}} & \shortstack{\texttt{valid}\\\texttt{@1}} & \shortstack{\texttt{secure-}\\\texttt{valid@1}} & \shortstack{\texttt{secure}\\\texttt{@1}} & \shortstack{\texttt{valid}\\\texttt{@1}} & \shortstack{\texttt{secure-}\\\texttt{valid@1}} & \shortstack{\texttt{secure}\\\texttt{@1}} & \shortstack{\texttt{valid}\\\texttt{@1}} & \shortstack{\texttt{secure-}\\\texttt{valid@1}} \\
\midrule
Llama & M01 & 55.17 & 66.17 & 43.17 & \underline{68.00} & 59.50 & 45.67 & 66.33 & 63.17 & \textbf{49.33} & \textbf{68.50} & 61.00 & \underline{48.17} & 67.33 & 57.67 & 43.33 \\
 & M02 & 73.17 & 63.67 & 48.50 & \underline{78.67} & 65.50 & \underline{56.50} & 70.50 & 59.00 & 50.00 & 77.50 & 61.00 & 56.33 & \textbf{82.33} & 64.00 & \textbf{61.17} \\
 & M03 & 67.67 & 50.83 & 40.33 & 68.17 & 43.83 & 37.33 & 69.00 & 39.33 & 35.50 & \underline{70.33} & 44.50 & \underline{39.67} & \textbf{74.33} & 48.33 & \textbf{44.17} \\
 & M04 & 63.67 & 64.50 & 44.50 & \underline{82.67} & 54.33 & 51.33 & 80.33 & 68.17 & \textbf{59.67} & 82.33 & 63.83 & \underline{58.50} & \textbf{83.33} & 60.17 & 56.83 \\
 & M05 & 58.00 & 84.17 & 46.83 & 48.83 & 83.83 & 38.33 & 58.83 & 84.67 & 46.50 & \underline{60.17} & 85.33 & \underline{49.67} & \textbf{62.83} & 83.17 & \textbf{50.67} \\
 & M06 & 61.17 & 82.50 & 50.50 & 74.00 & 76.67 & 58.67 & 77.33 & 81.67 & \underline{63.83} & \underline{77.67} & 81.50 & \textbf{65.67} & \textbf{78.33} & 81.83 & \textbf{65.67} \\
 & M07 & 64.67 & 69.50 & 49.00 & 67.83 & 65.67 & 48.00 & 74.00 & 68.17 & 53.17 & \underline{80.83} & 70.17 & \underline{59.83} & \textbf{85.33} & 70.50 & \textbf{61.50} \\
 & M08 & 36.67 & 73.83 & 27.67 & 42.50 & 70.17 & 29.67 & \textbf{51.67} & 72.00 & \textbf{37.83} & \underline{50.50} & 69.83 & \underline{35.17} & 48.83 & 71.83 & \underline{35.17} \\
 & M09 & 74.00 & 90.00 & 67.33 & \underline{94.17} & 85.50 & \underline{81.50} & 87.33 & 86.33 & 77.33 & 90.00 & 84.83 & 79.17 & \textbf{94.50} & 85.50 & \textbf{82.00} \\
 & M10 & 66.67 & 81.50 & 57.33 & \textbf{94.17} & 77.33 & \textbf{76.33} & 85.83 & 77.67 & 70.83 & 91.00 & 72.67 & 71.17 & \underline{93.67} & 72.00 & \underline{71.33} \\
\midrule
CodeGemma & M01 & 17.50 & 29.00 & 7.67 & \textbf{29.83} & 32.83 & \textbf{15.83} & 23.17 & 31.00 & 10.67 & \underline{25.50} & 29.17 & 11.00 & 25.00 & 35.00 & \underline{13.17} \\
 & M02 & 71.67 & 30.50 & 24.17 & \underline{74.00} & 35.33 & \textbf{25.83} & 73.67 & 30.67 & 23.50 & \textbf{74.17} & 28.83 & \underline{24.17} & 73.83 & 29.00 & 22.83 \\
 & M03 & 28.67 & 20.17 & 7.00 & 35.67 & 27.33 & 10.33 & \textbf{39.33} & 23.17 & \textbf{12.67} & 37.33 & 26.50 & \underline{12.33} & \underline{39.17} & 23.83 & 11.50 \\
 & M04 & 52.83 & 24.17 & 11.33 & 60.33 & 26.67 & 14.83 & 62.33 & 25.33 & 17.67 & \underline{65.83} & 27.67 & \textbf{20.17} & \textbf{67.50} & 25.83 & \underline{19.17} \\
 & M05 & 62.33 & 63.83 & 41.83 & 60.00 & 60.33 & 41.17 & 63.17 & 60.33 & 40.50 & \underline{64.67} & 64.50 & \underline{45.17} & \textbf{65.00} & 63.33 & \textbf{46.17} \\
 & M06 & 30.00 & 32.00 & 5.67 & 37.50 & 28.83 & 8.17 & \underline{41.83} & 25.83 & 9.00 & 40.50 & 29.50 & \underline{9.50} & \textbf{46.17} & 30.50 & \textbf{11.67} \\
 & M07 & 43.67 & 23.50 & 8.67 & 48.83 & 25.17 & 10.83 & 52.00 & 25.67 & 12.67 & \underline{56.00} & 26.50 & \underline{14.00} & \textbf{58.17} & 28.33 & \textbf{15.00} \\
 & M08 & 24.50 & 27.83 & 13.17 & 30.50 & 27.00 & 12.00 & \textbf{34.17} & 26.17 & \textbf{14.00} & 30.67 & 23.50 & 10.50 & \underline{31.33} & 31.17 & \underline{13.17} \\
 & M09 & 76.33 & 63.83 & 52.00 & \textbf{91.50} & 65.17 & \textbf{63.00} & 85.83 & 61.50 & 59.83 & 89.83 & 64.33 & \underline{61.83} & \underline{90.67} & 61.67 & 59.83 \\
 & M10 & 55.17 & 64.33 & 41.83 & \textbf{72.50} & 65.67 & \textbf{54.67} & 61.83 & 59.83 & 42.50 & \underline{70.00} & 63.17 & 50.83 & \underline{70.00} & 64.33 & \underline{52.83} \\
\midrule
Qwen & M01 & 29.17 & 67.17 & 24.83 & 39.50 & 63.67 & 29.83 & 40.83 & 64.00 & \underline{31.00} & \underline{43.67} & 63.67 & \underline{31.00} & \textbf{43.83} & 67.00 & \textbf{32.50} \\
 & M02 & 64.33 & 75.00 & 51.17 & 70.33 & 69.00 & 53.67 & 78.67 & 75.50 & 59.83 & \textbf{79.83} & 78.83 & \textbf{65.50} & \underline{79.33} & 75.33 & \underline{61.67} \\
 & M03 & 46.83 & 54.33 & 27.83 & \underline{53.17} & 50.00 & 29.33 & 51.83 & 53.33 & \underline{30.67} & \textbf{54.67} & 58.00 & \textbf{33.67} & 52.33 & 53.50 & 29.50 \\
 & M04 & 78.83 & 66.67 & 55.17 & 90.33 & 66.00 & 63.00 & 87.33 & 66.67 & 61.83 & \textbf{91.83} & 66.67 & \underline{64.67} & \underline{90.67} & 69.50 & \textbf{66.67} \\
 & M05 & 58.50 & 85.67 & 51.00 & 55.17 & 84.67 & 49.50 & 60.17 & 85.33 & 52.67 & \underline{62.33} & 86.67 & \textbf{54.50} & \textbf{63.00} & 85.67 & \underline{54.33} \\
 & M06 & 49.50 & 80.83 & 38.33 & \underline{70.17} & 74.00 & \textbf{52.67} & 66.83 & 74.00 & 46.67 & 69.17 & 76.17 & \underline{49.33} & \textbf{71.00} & 71.50 & 48.33 \\
 & M07 & 45.17 & 75.83 & 37.33 & 52.17 & 71.17 & 41.33 & 53.17 & 74.50 & 40.33 & \textbf{71.17} & 74.50 & \textbf{52.50} & \underline{68.00} & 74.00 & \underline{51.50} \\
 & M08 & 14.00 & 70.83 & 9.50 & \textbf{24.17} & 67.33 & 13.00 & 19.83 & 69.83 & \underline{13.83} & \underline{22.00} & 71.50 & \textbf{14.33} & 19.83 & 70.67 & 13.33 \\
 & M09 & 81.83 & 85.83 & 72.17 & 88.50 & 83.17 & 78.00 & \underline{91.33} & 83.67 & \underline{79.67} & \textbf{93.33} & 81.33 & \textbf{80.00} & 90.50 & 80.00 & 77.50 \\
 & M10 & 93.50 & 84.67 & 81.00 & \textbf{93.00} & 78.67 & \textbf{76.33} & 91.00 & 73.50 & 70.00 & \underline{91.50} & 73.67 & \underline{70.33} & 91.17 & 72.83 & 70.00 \\
\midrule
Qwen-Coder & M01 & 54.00 & 76.83 & 47.33 & \textbf{60.17} & 81.83 & \textbf{51.00} & 55.83 & 75.33 & 43.33 & \underline{58.50} & 71.83 & 43.33 & 57.83 & 76.17 & \underline{45.83} \\
 & M02 & 55.67 & 68.00 & 41.17 & 62.83 & 68.67 & 47.17 & 61.17 & 67.67 & 45.50 & \underline{63.67} & 67.83 & \underline{48.83} & \textbf{66.50} & 69.83 & \textbf{52.83} \\
 & M03 & 67.83 & 64.33 & 47.00 & 68.67 & 62.50 & \textbf{47.67} & 69.17 & 60.00 & 44.33 & \underline{70.50} & 59.50 & \underline{44.67} & \textbf{71.33} & 59.17 & 44.00 \\
 & M04 & 77.83 & 77.67 & 63.33 & 83.33 & 74.50 & 66.33 & \textbf{86.33} & 82.33 & \textbf{75.33} & \underline{85.83} & 79.50 & \underline{72.83} & 84.33 & 77.67 & 69.50 \\
 & M05 & 63.83 & 80.67 & 48.17 & 60.33 & 79.33 & 45.67 & 62.00 & 79.17 & 46.67 & \underline{63.83} & 81.33 & \underline{50.17} & \textbf{65.50} & 81.67 & \textbf{51.17} \\
 & M06 & 54.50 & 72.83 & 38.50 & 62.83 & 63.17 & 35.17 & \underline{65.33} & 67.00 & 39.67 & \textbf{68.67} & 67.67 & \underline{43.00} & 64.83 & 69.33 & \textbf{43.17} \\
 & M07 & 55.50 & 61.50 & 39.67 & 59.67 & 64.17 & 43.83 & \underline{63.00} & 61.17 & 41.50 & \textbf{66.50} & 61.17 & \underline{44.33} & \textbf{66.50} & 62.50 & \textbf{48.17} \\
 & M08 & 22.83 & 62.33 & 13.33 & 27.83 & 64.00 & \textbf{19.00} & \textbf{30.33} & 62.67 & \underline{18.33} & \underline{29.33} & 61.00 & 17.67 & 25.50 & 61.17 & 14.33 \\
 & M09 & 94.00 & 88.50 & 84.00 & 94.33 & 86.17 & 81.83 & \underline{94.83} & 87.00 & \underline{83.17} & 94.67 & 87.00 & \textbf{83.33} & \textbf{95.00} & 85.17 & 81.83 \\
 & M10 & 96.33 & 85.50 & 83.33 & \underline{97.67} & 82.50 & \textbf{81.67} & 97.00 & 76.50 & 76.17 & 95.50 & 75.50 & 74.67 & \textbf{97.83} & 78.83 & \underline{78.17} \\
\midrule
DeepSeek & M01 & 46.17 & 86.33 & 41.83 & \textbf{56.50} & 84.17 & \textbf{51.00} & 52.17 & 82.50 & 46.33 & 52.67 & 83.83 & 45.83 & \underline{53.67} & 83.33 & \underline{47.33} \\
 & M02 & 47.00 & 89.33 & 41.50 & 51.50 & 91.33 & 45.50 & \textbf{52.00} & 93.17 & \underline{47.67} & 50.67 & 93.00 & 47.00 & \underline{51.83} & 92.50 & \textbf{47.83} \\
 & M03 & 54.50 & 87.83 & 47.17 & 61.17 & 77.67 & 46.33 & 61.83 & 80.17 & 46.83 & \textbf{64.33} & 82.17 & \underline{52.33} & \underline{64.00} & 82.17 & \textbf{52.50} \\
 & M04 & 61.83 & 76.50 & 44.83 & \textbf{85.17} & 74.00 & \textbf{66.00} & 74.50 & 72.17 & 57.67 & \underline{79.33} & 74.33 & \underline{62.83} & 77.83 & 74.83 & 61.17 \\
 & M05 & 68.67 & 91.17 & 60.50 & 66.00 & 88.33 & 55.33 & \textbf{71.00} & 90.17 & \textbf{61.67} & \underline{70.67} & 88.67 & 59.83 & \underline{70.67} & 88.83 & \underline{61.00} \\
 & M06 & 68.83 & 91.83 & 65.83 & \textbf{65.17} & 89.33 & \textbf{61.67} & 58.17 & 90.00 & 55.83 & 61.17 & 86.50 & 56.67 & \underline{62.83} & 87.00 & \underline{57.50} \\
 & M07 & 79.83 & 88.33 & 72.67 & 82.17 & 88.67 & \underline{74.17} & \underline{83.67} & 83.67 & 72.17 & \textbf{86.67} & 86.33 & \textbf{77.67} & 83.00 & 82.50 & 70.50 \\
 & M08 & 25.33 & 93.00 & 24.17 & \textbf{48.33} & 90.50 & \textbf{45.67} & \underline{45.50} & 91.00 & \underline{42.50} & 42.00 & 91.50 & 39.17 & 43.17 & 91.33 & 39.83 \\
 & M09 & 89.33 & 97.50 & 87.83 & \textbf{97.00} & 95.33 & \textbf{92.50} & 94.17 & 94.17 & 89.17 & 94.50 & 92.33 & 88.33 & \underline{96.00} & 94.33 & \underline{91.00} \\
 & M10 & 77.17 & 89.33 & 72.00 & 77.33 & 84.67 & 69.83 & \underline{83.17} & 92.17 & \textbf{80.00} & 82.67 & 88.00 & \underline{76.50} & \textbf{84.17} & 92.33 & \textbf{80.00} \\
\bottomrule
\end{tabular*}
\caption{
Single-generation success rates over 600 outputs per model and condition for each misuse category.
Bold and underlining mark the highest and second-highest \texttt{secure@1} and \texttt{secure-valid@1} among the four directions at $+1\sigma$.}
\label{tab:app-misuse-method-comparison}
\end{table*}

\section{Generation Evaluation Details and Subgroup Results}
\label{app:generation_evaluation_details_and_subgroup_results}

We report the aggregate generation results in Section~\ref{sec:steering_code_generation_with_shared_repairs}.
Here we provide paired statistical comparisons and breakdown results by language and misuse category to quantify uncertainty and examine how steering effects vary across languages and cryptographic functionalities. 

\subsection{Paired Comparisons} \label{app:paired_comparisons}
We pair outputs across conditions by model, language--misuse category cell, prompt, and generation seed. 

To estimate uncertainty in the aggregate difference, we compute 95\% confidence intervals using 20,000 paired hierarchical bootstrap replicates. 
Within each of 40 cells, we resample five prompts and then 30 generation attempts per prompt with replacement, preserving paired indices across conditions~\cite{hesterberg_2011_bootstrap}.
Then we average the resulting cell rates with equal weight. 
The 95\% confidence interval is defined by the 2.5th and 97.5th percentiles of the replicated rate differences. 

We further conduct significance tests that treat each prompt as a cluster of 30 attempts. 
We average the paired differences within each prompt, then apply a two-sided sign-flip test with 100,000 replicates~\cite{dror_2018_hitchhikers}.
Each replicate independently keeps or reverses the sign of each prompt mean. 
The test compares the absolute aggregate difference with the resulting distribution. 
For each model and metric, the unsteered baseline and four steering methods at $+1\sigma$ define a family of $\binom{5}{2}=10$ pairwise comparisons. 
We apply Holm corrections to this comparison family~\cite{holm_1979_simple}, and use an adjusted significance threshold of 0.05.
Table~\ref{tab:app-key-contrasts} supplements Table~\ref{tab:generation-main} by reporting four of the ten comparisons, contrasting \method{} with the unsteered baseline, Cell DiffMean, Pooled DiffMean, and Normalized DiffMean. 
The stars denote Holm-adjusted significance.

\subsection{Subgroup Results} \label{app:subgroup_results}
Tables~\ref{tab:app-language-method-comparison} and~\ref{tab:app-misuse-method-comparison} report unsteered baseline and $+1\sigma$ steering results broken down by language and misuse category.
These results show how steering effects and method rankings vary across tasks. 
\method{} ranks first or second on \texttt{secure-valid@1} in 16 of 20 model--language groups and 33 of 50 model--category groups.

The misuse category M05 gives a recurring example of the benefit of the shared direction for steering. 
Cell DiffMean lowers both \texttt{secure@1} and \texttt{secure-valid@1} relative to the baseline in all five models, but \method{} increases both in all models. 
For this misuse category, the shared direction provides more effective generation intervention than the mean direction fitted specifically to the target misuse category.


\begin{table*}[t]
\centering
\footnotesize
\setlength{\tabcolsep}{4pt}
\begin{tabular}{@{}ll*{3}{r@{\hspace{0.25em}}l}@{}}
\toprule
Model & Comparator &
\multicolumn{2}{c}{$\Delta$\texttt{secure@1}} &
\multicolumn{2}{c}{$\Delta$\texttt{valid@1}} &
\multicolumn{2}{c}{$\Delta$\texttt{secure-valid@1}} \\
\midrule

\multirow{2}{*}{Llama} & SCS-Code
& $+1.63$ & $[-0.17,3.43]$
& $-3.12$ & $[-5.05,-1.18]^*$
& $-1.28$ & $[-3.38,0.80]$ \\
& Surgical Repair
& $+10.52$ & $[8.42,12.65]^*$
& $-1.37$ & $[-3.42,0.68]$
& $+6.57$ & $[4.27,8.87]^*$ \\
\midrule

\multirow{2}{*}{CodeGemma} & SCS-Code
& $+5.18$ & $[3.02,7.32]^*$
& $-1.05$ & $[-3.13,1.05]$
& $+1.52$ & $[-0.38,3.42]$ \\
& Surgical Repair
& $+5.48$ & $[3.40,7.55]^*$
& $-0.28$ & $[-2.42,1.85]$
& $+2.73$ & $[0.90,4.55]^*$ \\
\midrule

\multirow{2}{*}{Qwen} & SCS-Code
& $+3.80$ & $[1.57,6.03]^*$
& $-1.93$ & $[-4.33,0.35]$
& $+0.48$ & $[-2.15,3.00]$ \\
& Surgical Repair
& $+6.67$ & $[4.62,8.75]^*$
& $+0.58$ & $[-1.53,2.68]$
& $+5.42$ & $[3.10,7.77]^*$ \\
\midrule

\multirow{2}{*}{Qwen-Coder} & SCS-Code
& $+4.12$ & $[2.53,5.72]^*$
& $-2.15$ & $[-3.92,-0.40]$
& $+1.47$ & $[-0.47,3.38]$ \\
& Surgical Repair
& $+5.08$ & $[3.47,6.70]^*$
& $-2.12$ & $[-3.98,-0.25]$
& $+1.95$ & $[0.10,3.82]$ \\
\midrule

\multirow{2}{*}{DeepSeek} & SCS-Code
& $+6.57$ & $[4.20,8.97]^*$
& $+0.93$ & $[-1.15,3.03]$
& $+5.88$ & $[3.18,8.63]^*$ \\
& Surgical Repair
& $+5.32$ & $[2.20,8.52]^*$
& $-1.60$ & $[-3.38,0.18]$
& $+4.23$ & $[1.03,7.52]$ \\
\bottomrule
\end{tabular}
\caption{\method{} minus each comparator at $+1\sigma$.
Differences are in percentage points, with 95\% confidence intervals
over 6,000 paired attempts.
$^*$ denotes Holm-adjusted $p<0.05$.}
\label{tab:app-literature-contrasts}
\end{table*}

\section{Comparison with SCS-Code and Surgical Repair} \label{app:comparison_with_scscode_and_surgical_repair}
Section~\ref{sec:comparing_with_other_security_steering_methods} compares three methods of constructing a security direction. 
The experiment uses the same checkpoint, tokenizer, 200 prompts, and 30 attempt seeds across methods within a model.
We use the common layer, scale, and activation update in Equation~\ref{eq:generation-intervention}, and replace the original method's layer and strength searches, to separate the influecen of the direction-construction method. 


SCS-Code \cite{wendlinger_2026_security} uses all 48,000 matched program pairs in our resource. 
Each input contains the two programs as two options and asks the model to choose one snippet.
We use the original code-choice template with the language name and code-fence tag replaced as appropriate, which is shown below.
Following their original design, a shuffle assigns the secure answer to A and B 24,000 times each.
We subtract the misuse-selecting token activation from the secure-selecting token activation at the selected layer to obtain a direction from each pair. 
\begin{lstlisting}[style=appendixprompt]
Hello, I am a researcher on <language name> programming code. I will give you two <language name> code snippets in the following, marked with (A) and (B). You need to choose one code snippet. Here are the snippets:
(A) ```<language tag>
<code A>
```

(B) ```<language tag>
<code B>
```

Which snippet do you choose, (A) or (B)?
\end{lstlisting}

For Surgical Repair~\cite{sandoval_2026_surgical}, we use 144 plain-text instruction pairs without a chat template. 
Each instruction pair requests the same cryptographic functionality with contrasting secure and misuse implementation choices. 
For each pair, we obtain a direction by subtracting the activation at the final token of the insecure instruction from that of the secure instruction at the common selected layer.

For both methods, we average the secure-minus-misuse activation differences with equal pair weights and normalize the resulting vector before applying the common intervention scale. 
We compare the generation outcomes of the three direction-construction methods using paired statistics.
Table~\ref{tab:app-literature-contrasts} reports differences over 6,000 paired outputs per model, with a common intervention strength of $+1\sigma$. 
We use the same significance tests as in Appendix~\ref{app:generation_evaluation_details_and_subgroup_results} and apply Holm corrections to all six pairwise comparisons among the unsteered baseline and three methods.

\section{Evaluation and Output Audits}
\label{app:evaluation_and_output_audits}
We evaluate structural validity and target security in model-generated outputs and audit the evaluators.
The three metrics in Section~\ref{sec:intervention_and_evaluation} use the single-sample case of the HumanEval \texttt{pass@k} estimator~\cite{chen_2021_evaluating}.
For $c$ successful samples among $n$ attempts, the estimator is
\begin{equation}
 \widehat{\mathrm{pass@}k}
 =1-\frac{\binom{n-c}{k}}{\binom nk}.
 \label{eq:pass-at-k}
\end{equation}
We use $n=30$ attempts per request, use $k=1$ for the reported results, and use the secure label, structural validity, or both on the same output as the criterion. 

\begin{table*}[tbp]
\centering
\footnotesize
\setlength{\tabcolsep}{2pt}
\begin{tabularx}{\textwidth}{@{}l>{\raggedright\arraybackslash}X>{\raggedright\arraybackslash}X@{}}
\toprule
ID  & Misuse & Secure use\\
\midrule
M01 & Cryptographic key from a fixed string, fixed byte array, or general-purpose PRNG & Cryptographic key generated by a CSPRNG or derived by a KDF \\
M02 & AES with ECB mode & AES encryption with other secure modes, such as CBC, CTR, GCM, CCM, OFB, CFB \\
M03 & AES CBC encryption with IV from fixed bytes, a fixed string, or predictable randomness & AES CBC encryption with IV generated by a CSPRNG \\
M04 & RSA key size $<2{,}048$ bits & RSA key size $\geq 2{,}048$ bits \\
M05 & General-purpose PRNG output or a predictable generator seed & CSPRNG output \\
M06 & Fixed salt, or predictably reused salt & Fresh random salt \\
M07 & PBKDF2/PBE iteration count $<1{,}000$ & PBKDF2/PBE iteration count $\geq 1{,}000$ \\
M08 & Hardcoded password or password derived from constants or predictable randomness & Password from environment, user input or a CSPRNG \\
M09 & Broken hash functions: MD2, MD4, MD5, or SHA-1 & Other secure hash functions, such as SHA-224/256/384/512, SHA3-224/256/384/512 \\
M10 & Compromised MAC algorithms: HMAC-MD5, HMAC-SHA-1 & Other secure MAC algorithms, such as HMAC with SHA-224/256/384/512 or SHA3-224/256/384/512  \\
\bottomrule
\end{tabularx}
\caption{Security rules for each misuse category. CSPRNG denotes a cryptographically secure pseudorandom number generator.}
\label{tab:app-security-policies}
\end{table*}

\paragraph{Structural Validity}
\label{app:structural_validity}

Structural validity checks whether the output contains a coherent implementation of the requested cryptographic functionality. 
It requires the following six checks to pass:
\begin{enumerate}[noitemsep]
 \item Parses or compiles.
 \item Uses resolvable imports.
 \item Uses supported APIs for the cryptographic functionality.
 \item Uses the API operations required by that cryptographic functionality.
 \item Uses compatible API signatures and argument roles.
 \item Has a connected sequence of operations to complete the cryptographic functionality.
\end{enumerate}
A program is structurally valid only when all six checks complete successfully. 
We choose structural validity to evaluate the utility of the generated programs because they have no common entry point or input/output interface, which makes execution-based functional testing not feasible. 
We conduct an exploratory AI-assisted review of 500 outputs to assess the feasibility of functional testing. 
Only 76 of them are suitable for reusable adapters for execution-based functional testing. 
This supports our choice of using structural validity to evaluate the utility of generated programs. 

\paragraph{Security of API Usage} \label{app:security_of_api_usage}

We assess the security of the misuse category targeted by each task request.
We follow the security policy for benchmark evaluation, and Table~\ref{tab:app-security-policies} lists the parameters, algorithm choices or thresholds used to classify generated programs.

The evaluator combines evidence from the target API uses into a program-level label, because there might be multiple target API uses in a program.
When all target API use is secure, we label a program as \emph{secure}.
When there is misuse API uses but no secure use, we label a program as \emph{misuse}.
When both secure and misuse API use exist in the same program, we label it as \emph{mixed}.
Otherwise, we label the program as \emph{unknown}, such as in cases where the security of API usage depends on user input.
We report \texttt{secure@1} and \texttt{secure-valid@1} as the main outcomes because our main goal is to increase the proportion of outputs that are secure.

\paragraph{Evaluator Checks} \label{app:evaluator_checks}
We implement evaluators tailored to these structural and security criteria.
We test the quality of evaluators by a blinded review. 
Within each language--misuse category cell for every model, we randomly selected one sample from 150 generations in that cell for the unsteered baseline, and chose the sample from Normalized DiffMean and \method{} with $+1\sigma$ steering that share the same prompt and generation random seed with that sample. 
This gives a set of 600 outputs. 
The first author manually reviewed these generations and assigned labels for security and validity. 
Among the 600 samples, 16 of them have disagreements between manual and evaluator labels for security, and 9 of them have disagreements for validity. 
This gives an overall agreement of 97.33\% for security labels, and 98.50\% for structural validity. 
Replacing evaluator labels with human judgments within this sample preserves the signs of the comparison of \texttt{secure@1}, \texttt{valid@1}, \texttt{secure-valid@1} among unsteered baseline, Normalized DiffMean, and \method{}, with changes of at most 1.50 percentage points. 
This suggests that the evaluator results are reliable and not likely to affect the overall results.

\section{Reproducibility and Computational Budget} \label{app:reproducibility_and_computational_budget}

\begin{table}[tbp]
\centering
\footnotesize
\setlength{\tabcolsep}{0.8pt}
\sisetup{
  mode=text,
  table-number-alignment=center
}
\begin{tabular}{@{}l
  S[table-format=1.1]
  S[table-format=1.1]
  S[table-format=3.1]
  S[table-format=3.1]
@{}}
\toprule
Model &
{\shortstack{Activation\\extraction}} &
{\shortstack{Layer\\probes}} &
{\shortstack{Full-source\\generation}} &
{\shortstack{Source-exclusion\\generation}} \\
\midrule
Llama & 1.8 & 1.1 & 124.0 & 65.1 \\
CodeGemma & 2.5 & 0.8 & 104.5  & 56.1\\
Qwen & 1.8 & 1.1 & 60.6 & 33.0 \\
Qwen-Coder & 1.9 & 1.3 & 72.0 & 39.3 \\
DeepSeek & 3.0 & 1.0 & 194.7 & 103.2\\
\bottomrule
\end{tabular}
\caption{Approximate elapsed hours by model and stage. Full source generation includes the unsteered baseline.}
\label{tab:app-compute-time}
\end{table}

Our code and data are available at
\url{https://github.com/tianhe-lu/CrypSteer}.
We use one NVIDIA GeForce RTX 4090 for experiments with Llama, CodeGemma, Qwen and Qwen-Coder models, and use one NVIDIA A100-SXM4-80GB GPU for DeepSeek.  
Table~\ref{tab:app-compute-time} reports the approximate elapsed time for the main GPU-based stages of the study.
We used fixed steering strength settings without a hyperparameter search process.

\end{document}